\documentclass{emulateapj}
\usepackage{epsfig}
\begin{document}
\def\eg{{\it e.g.}}
\def\ie{{\it i.e.}}
\newcommand{\tnm}[1]{\tablenotemark{#1}}
\title{The Star Formation History of the Large Magellanic Cloud}

\author{Jason Harris}
\affil{National Optical Astronomy Observatory, 950 North Cherry Ave., Tucson, AZ, 85719}
\and
\author{Dennis Zaritsky}
\affil{Steward Observatory, 933 North Cherry Ave., Tucson, AZ, 85721}

\begin{abstract}
We present the first-ever global, spatially-resolved reconstruction of
the star formation history (SFH) of the Large Magellanic Cloud (LMC),
based on the application of our StarFISH analysis software to the
multiband photometry of twenty million of its stars from the
Magellanic Clouds Photometric Survey.  The general outlines of our
results are consistent with previously published results: following an
initial burst of star formation, there was a quiescent epoch from
approximately 12 to 5 Gyr ago.  Star formation then resumed and has
proceeded until the current time at an average rate of roughly
0.2~$M_\odot$~yr$^{-1}$, with temporal variations at the factor-of-two
level. The re-ignition of star formation about 5 Gyr ago, in both the
LMC and SMC, is suggestive of a dramatic event at that time in the
Magellanic system.  Among the global variations in the recent star formation
rate are peaks at roughly  2~Gyr, 500~Myr, 100~Myr and 12~Myr.  The peaks at
500~Myr and 2~Gyr are nearly coincident with similar peaks in the SFH
of the Small Magellanic Cloud, suggesting a joint history for these
galaxies extending back at least several Gyr.  The chemical enrichment
history recovered from our StarFISH analysis is in broad agreement
with that inferred from the LMC's star cluster population, although
our constraints on the ancient chemical enrichment history are weak.
We conclude from the concordance between the star formation and
chemical enrichment histories of the field and cluster populations
that the field and cluster star formation modes are tightly coupled.
\end{abstract}


\section{Introduction}\label{sec:intro}

The LMC is an important target in the effort to understand the stellar
populations of galaxies, due to its close proximity, favorable viewing
angle, and ongoing star-formation activity.  Detailed examination of
the LMC's stellar content began in the 1960s with much of it focused
on the star cluster population.  \cite{hod60, hod61} surveyed populous
clusters in the LMC, finding that while 35 of the clusters are similar
in age to the population of globular clusters in the Milky Way, the
LMC contains 23 populous clusters that are clearly much younger than
any Milky Way globular cluster.  This observation suggested that the
star formation history (SFH) of the LMC might be fundamentally
different than that of the Milky Way.  Age-dating of LMC clusters via
isochrone fitting to color-magnitude diagrams (CMDs) commenced with
the dawn of the CCD era, and by 1989 over 100 LMC clusters had been so
analyzed \citep{sp89}.  The age-metallicity relation (AMR) of the
LMC's star clusters contains some perplexing features, such as the
lack of clusters between 5 and 12~Gyr old \citep{dc91, osm96, pia02},
and a wide metallicity dispersion among the young ($t<3$~Gyr) clusters
\citep[\eg, ][]{gei03}.  These features provide clues to the history
of star formation in the LMC, but their interpretation has proved
difficult.

As useful as star clusters are, a complete understanding of the LMC's
star formation history (SFH) requires analysis of its entire stellar
population.  Early work on the LMC's field population included the
analysis of the spatial distribution of Cepheid variables in the LMC,
grouped by age \citep{pg72}.  She found abundant evidence for a
complex, spatially-variable SFH, albeit over a relatively limited age
interval of $10^7$--$10^8$~yr.  The Cepheid analysis was updated by
the MACHO project \citep{alc99}, who found a burst of star formation
115~Myr ago.

Understanding star formation in the LMC across its entire history
requires disentangling the complex admixture of its field stellar
populations, a task that was long regarded as nearly impossible
\citep[\eg, ][p. 2]{sch58}.  However, by the 1980s, our understanding
of stellar evolution and the stellar initial mass function (IMF)
coupled with the improvements in computing and detector technology
made such an analysis possible.  The key to disentangling the complex
mix of stellar populations in the LMC lies in the quantitative
comparison of observed CMDs to synthetic CMDs built by populating
theoretical isochrones according to an adopted IMF and an input model
star formation history.  Some of the earliest synthetic-CMD efforts
employed the ``R method'', which is a statistical comparison of
observed and synthetic CMDs using the main sequence luminosity
function and ratios of star counts in age-sensitive regions of the
CMD, such as the red giant branch and the subgiant region
\citep{ber92}.

All modern synthetic CMD methods are essentially generalizations of
the R method: instead of a one-dimensional luminosity function and
star-count ratios in a few select regions of the CMD, the full
N-dimensional distribution of stars in one or more CMDs is employed in
the analysis.  Much of the recent work on reconstructing the SFH of
the LMC has utilized data from the Hubble Space Telescope (HST), which
provides CMDs that reach well below the ancient main sequence turn-off
\citep[MSTO; ][]{glg96, geh98, hol99, ols99, sh02}, even in the
crowded bar regions.  Such depth is difficult to achieve with
ground-based observations due to the limited resolution, especially in
the crowded regions of the LMC bar.  Reaching below the ancient MSTO
is important not only for constraining the ancient SFH, but also for
measuring the low-mass stellar IMF, which provides the critical
conversion factor for ground-based stellar photometry between the
number of bright stars observed and the total stellar mass present.

Among the HST studies targeting the bar, there is a remarkable
consensus regarding the general outlines of the LMC's early SFH.  The
``Age Gap'' seen among the LMC's star clusters is also evident in the
field populations of the bar: following an initial burst of star
formation, the LMC experienced a long quiescent epoch that lasted
several~Gyr.  Then, 4 or 5~Gyr ago, star formation resumed and
continues to this day.  The question of the Age Gap among field
populations in the LMC disk is more ambiguous.  The study of some areas of the
disk seem to demand a quiescent epoch \citep[cf.][]{dol00}, while others are more
consistent with continuous star formation throughout the LMC's early
history \citep[cf.][]{geh98}.

Despite their exquisite depth, the collected
SFH solutions based on HST data are necessarily incomplete, simply
because the HST field of view is vanishingly small compared to the
solid angle subtended by the LMC: a full-coverage survey with ACS
would require roughly 50,000 pointings.  While we expect the older
stellar populations to be well-mixed throughout the LMC, its stellar
disk has a solid-body rotation curve out to roughly 4~kpc
\citep{vdm02} and patterns imprinted by a spatially-variable star
formation rate could persist in the LMC for several rotational periods
($> 1$ Gyr).  Until now, a truly global analysis of the LMC's SFH over
its entire history has not been attempted.
 
We present a spatially-resolved solution for the 
global SFH of the LMC, based on a photometric catalog of over 20 
million LMC stars.  In Section~\ref{sec:data}, we briefly review 
our photometric catalog.  In Section~\ref{sec:starfish}, we review 
our StarFISH synthetic CMD algorithm, and present some details on its 
application to the LMC data.  We present our spatially-resolved map 
of the LMC's SFH in Section~\ref{sec:sfhmap}, and discuss some of its 
implications in Section~\ref{sec:discuss}.  The paper is summarized 
in Section~\ref{sec:summary}.

\section{The Magellanic Clouds Photometric Survey}\label{sec:data}

\begin{figure}[htbp]
\epsscale{0.7}
\plotone{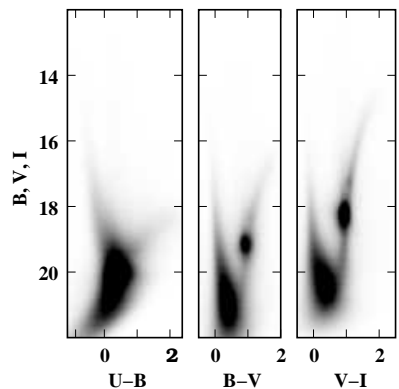}
\caption{A ``CMD triptych'' illustrating our $UBVI$ photometry of
  $\sim$20 million LMC stars.  Each CMD panel is a Hess diagram whose
  pixel values are proportional to the number of stars in each pixel.
  The mixed populations evident in these CMDs represent a ``fossil
  record'' of a complex star-formation history in this
  galaxy. \label{fig:cmds} }
\end{figure}

The Magellanic Clouds Photometric Survey \citep[MCPS; ][]{zar97} was
undertaken at the Las Campanas Observatory's 1-meter Swope Telescope,
between 1995 and 2000.  It was a drift-scan survey using the Great
Circle Camera \citep{zar96}, covering $4^\circ\times4.6^\circ$ in the
Small Magellanic Cloud \citep{zar02}, and $8.5^\circ\times7.5^\circ$
in the LMC \citep{zar04}.  We obtained images with $U$, $B$, $V$ and
$I$ filters, reaching a limiting magnitude between $V=20$ and
$V=21$~mag, depending on the local degree of crowding in the images.
For details on the data reduction, construction of the photometric
catalog, and quality assurance tests, see \cite{zar02}.  The final LMC
photometric catalog contains over 24~million objects, the vast
majority of which are stars in the LMC.  The photometry is presented
in Figure~\ref{fig:cmds} in the form of a ``CMD triptych'' (three Hess
diagrams showing $U-B$ vs. $B$, $B-V$ vs. $V$, and $V-I$ vs. $I$).

We use both the $B-V$ and $V-I$ CMD planes in determining the best-fit
SFH.  StarFISH can easily be configured to operate simultaneously on
any number of photometric planes (CMDs or two-color diagrams).
Throughout this paper, when we are specifically discussing the present
analysis, we will use the term ``CMD pair'' to describe the StarFISH
analysis unit.  However, when discussing StarFISH in general, we may
use the more generic term ``CMD''.

%CMD triptych for entire LMC population

\section{Deriving the LMC's Star Formation History with StarFISH}\label{sec:starfish}

StarFISH \citep{hz01} is the software package that we created
specifically for the synthetic-CMD SFH analysis of our MCPS data.
However, we developed StarFISH with generality in mind, and to date it
remains one of the few synthetic CMD packages that is publicly
available for use by the astronomical community\footnotemark.  We
have already applied StarFISH to the MCPS catalog of the Small
Magellanic Cloud \citep[SMC, ][]{hz04}, and the present analysis will
be similar to that work; all important differences are noted in
this Section.

\footnotetext{http://ngala.as.arizona.edu/mcsurvey/}

StarFISH works by constructing a library of synthetic CMDs, each of
which represents the predicted distribution in color and brightness
for a stellar population of a particular age and metallicity.  The
library spans appropriate ranges in age and metallicity, so that any
composite population can be generated as a linear sum of all the
synthetic CMDs.  The SFH of an observed stellar population is
determined by minimizing the differences between the observed and
model CMDs, via adjustment of the amplitudes.  The collection of
best-fit amplitude values -- the proportional contribution of model
stars of each age and metallicity -- represents the best-fit SFH of
the observed stellar population.

\subsection{The Isochrone Set}\label{sec:isoc}

StarFISH can operate with any set of isochrones.  We choose the latest
isochrones from the Padua group \citep{gir02}, because they provide a
uniform set covering the relevant range of ages and metallicities, and
include important post-red giant branch phases of stellar evolution
(in particular, the red clump and asymptotic giant branch).

To prepare the isochrones for synthetic CMD construction, we first
interpolate along each isochrone to increase the photometric
resolution such that the photometric distance between adjacent points
does not exceed 0.005 mag.  This procedure ensures that when we
populate the isochrones with artificial stars, there are no visible
irregularities.  Next, we determine the ``occupation probability,''
the relative probability of finding a star between points $i$ and
$(i+1)$ along the isochrone for a given IMF.  For a simple power-law
IMF, the occupation probability is given by: $P = (M(i+1)^\Gamma -
M(i)^\Gamma)/\Gamma$ where $M(i)$ is the {\it initial} mass of a given
isochronal point, and $\Gamma$ is the power-law index (\ie,
$\Gamma=-1.35$ for a Salpeter IMF).

Finally, we adjust the age and metallicity resolution provided by the
base isochrones to be appropriate for the observed data.  For example,
deep {\it HST} observations allow for much finer age resolution than
shallow, ground-based data.  This adjustment of the isochrone set
involves subjective judgment by the user, but our rule of thumb is
that we want a set of synthetic CMDs that each represent a unique
distribution in the CMD plane.  In other words, if a pair of
isochrones is indistinguishable after adding photometric errors, then
they should be combined into a single ``group'' that will be treated
as one entity by the fitter.  Similarly, if two adjacent isochrones
have a detectable gap between them after adding photometric errors,
then we use interpolation to insert an intermediate isochrone between
them.  StarFISH provides mechanisms for both grouping and
interpolating to optimize the age and metallicity resolution.  We find
that the Padua isochrones provide a higher age resolution than our
ground-based data warrant.  Therefore, we join adjacent isochrones
into age bins that span 0.2 dex, for log(Age) $>$ 8.0.
For ages younger than 100~Myr, we increase the age interval to
log(Age)=0.3 dex, resulting in bins with log(Age)=7.7, 7.4, 7.1, and
6.8.  The wider age bins are necessitated by the steepness of the
stellar IMF, which results in poorly-populated upper main sequences
and supergiant sequences.  
We begun
the analysis with isochrones of three metallicities (Z=0.001, Z=0.004,
and Z=0.008), each of which is provided in the Padua set of
isochrones.  It quickly became apparent that our best-fit solutions
would benefit from an additional metallicity bin, so we interpolated
between the Z=0.001 and Z=0.004 isochrones to generate a fourth set of
isochrones with Z=0.0025.  The four metallicity bins are
only used for the subset of isochrones with $\log(Age) \ge 8.0$.
Providing multiple metallicity bins for younger ages proved to be
counter-productive because the photometry of such stars is insensitive
to metallicity, and so allowing metallicity variations for these ages
only serves to inflate the fit uncertainties.

\subsection{Interstellar Extinction}\label{sec:dust}

When attempting to model stellar populations in a galaxy's diffuse 
field, it is important to match not only the mean extinction, but 
also the distribution of extinction values, as a function of age. 
Young stellar populations tend to have 
extinction values that are both larger on average and more widely 
distributed, than older stars.  Indeed, \cite{zar04} found that in the LMC
the mean extinction among hot stars was four times larger than among 
cool stars, and that the dispersion in the hot-star extinction 
distributions tended to be larger.  This difference likely arises 
because hot stars, which are typically younger than cool stars 
tend to be observed in the dustier environments in which 
stars form.  In contrast, older stars have had time to drift away from their 
birthplaces, and so 
occupy more typical, lower-extinction environments.

% extinction distros from selected regions
\begin{figure}[htbp]
\plotone{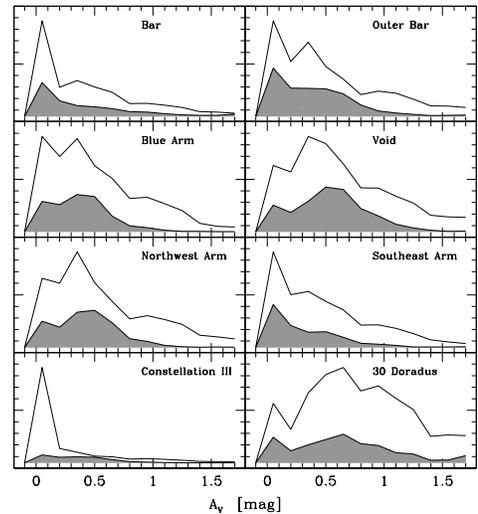}
\caption{Extinction distributions for both hot (unshaded line) and
  cool (shaded line) star samples, for eight small subregions sampling
  various substructures in the LMC, as labeled in each panel.  The
  plot illustrates the spatial variability in the extinction
  properties, as well as the wide distribution of extinction values
  within many of the regions. Derived negative extinctions arise due
  to photometric errors \cite[see][for more details]{zar04} and
  such values are set to $A_V = 0$. \label{fig:extdist} }
\end{figure}
                                                   
To address these complications, we
use distributions of extinction values that are either derived
directly from the stars in the region being analyzed, or from stars in
a different region with similar extinction properties. We refer the reader
to \citep{zar99, zar04} for details.
In Figure~\ref{fig:extdist}, we show some sample extinction
distributions from different locations in the LMC, illustrating the
degree to which extinction varies in this galaxy.  While the extinction
measurement for an individual star has significant uncertainty \citep[0.1 to 0.2
mag;][]{zar99}, the distribution is derived from many stars in each region.
In generating
synthetic CMD libraries for determining the SFH for each of these
regions, we draw extinction values randomly from the region's own
measured extinction distribution (assigning values from the hot-star
distribution to synthetic stars younger than 10~Myr, cool-star values
to synthetic stars older than 1~Gyr, and a linear interpolation
between the hot and cool star samples for synthetic stars with ages
between 10~Myr and 1~Gyr).  Thus, composite models constructed from
the synthetic CMD library will accurately reflect the full range of
the region's extinction properties, including its mean value,
dispersion, and distinct properties for its young and old components.
We are able to use a given synthetic CMD library as
the model for multiple regions, provided that the regions all share
similar crowding and extinction properties.  We do this 
because the generation of a synthetic CMD library is
computationally expensive.

\subsection{Photometric Errors}\label{sec:photerrs}

Before attempting to compare synthetic CMDs generated by StarFISH to
observational
CMDs, it is critical that the synthetic CMDs incorporate a
realistic photometric error model that include random and systematic
errors, and incompleteness fractions as a function of position in the
CMDs.  We perform artificial stars tests (ASTs) on the target images
to construct robust, empirical error models that accurately reproduce
the photometric properties of the real data.  ASTs are especially
important when the photometric errors are dominated by crowding,
rather than photon statistics, as is the case for these data.

To perform such an AST, we inject
artificial stars 
into the $B$, $V$ and $I$ images using
the world coordinate system (right ascension and declination), which
ensures that each artificial star appears in the same location with
respect to the real stars in the field in all images.
The artificial stars are placed in the image on a uniform grid,
separated by ten times the radius of the point-spread function in $V$.
This procedure ensures that each artificial star samples the crowding conditions
of the original data image, with no interference from other artificial
stars.  We assign a  $V$ magnitude drawn from an
exponential probability envelope that increases toward fainter
magnitudes
to better sample the faint end where errors and
incompleteness change rapidly.  $B$ and $I$ magnitudes are then
assigned by drawing $B-V$ and $V-I$ colors randomly over the range
covered by the real stellar populations.  
We perform several hundred AST
iterations on each frame, randomly displacing the grid positions for
each iteration, and reassigning the input photometry in each band.

We analyze the images with the artificial stars using the same DAOPhot photometry
pipeline as previously,
with the exception that we adopt the point-spread function model from the
original analysis, rather than re-deriving it.  We then 
record the $m_{input} - m_{recovered}$ difference in each band
for the artificial stars.  Artificial
stars that are not detected by the pipeline analysis are 
used to determine the completeness.  
The ensemble of photometric
$\Delta m$ and completeness data provides the empirical error model
we use to generate the synthetic CMDs.

Ideally, we would perform an independent AST for each 
LMC region. However, because the ASTs take
several days to complete even on modern multicore computers, it is not
practical to perform thousands of ASTs.  Instead, we ran ASTs on the
17 LMC subregions listed in Table~\ref{tab:ast} that are carefully selected to
be representative of the range of crowding conditions.

%Table of AST regions
\begin{deluxetable}{rrrr}
\tablecaption{The Artificial Stars Tests \label{tab:ast}}
\tablewidth{0pt}
\tablehead{ 
  \colhead{Subregion\tnm{a}} & 
  \colhead{$N_\star/\Box\arcmin$} & 
  \colhead{$\langle A_V(hot) \rangle$ [mag]} & 
  \colhead{$\langle A_v(cool) \rangle$ [mag]} 
}
\startdata
DP &  47.4 & 0.7  & 0.4 \\
EP &  69.4 & 0.7  & 0.4 \\
CF &  70.9 & 0.7  & 0.6 \\
JC &  77.7 & 0.6  & 0.5 \\
GC &  79.0 & 0.5  & 0.5 \\
FP &  94.3 & 0.5  & 0.4 \\
EF & 105.3 & 0.6  & 0.5 \\
KC & 105.9 & 0.7  & 0.5 \\
LC & 125.7 & 0.6  & 0.5 \\
PN & 131.2 & 0.4  & 0.4 \\
MN & 142.9 & 0.4  & 0.4 \\
ON & 145.9 & 0.25 & 0.4 \\
NN & 160.7 & 0.2  & 0.3 \\
KF & 174.6 & 0.5  & 0.4 \\
HF & 181.0 & 0.5  & 0.4 \\
IF & 186.7 & 0.5  & 0.4 \\
OF & 221.1 & 0.4  & 0.3 \\
PD & 174.6 & 0.7  & 0.5 \\
PI & 142.9 & 0.8  & 0.7 \\
PM & 131.2 & 0.5  & 0.5 \\
\enddata
\tablenotetext{a}{Each two-letter code indicates the region's position
in our gridding of the LMC (see Figure~\ref{fig:datagrid}).}
\label{tab:ast}
\end{deluxetable}                                            

% CMDs from select regions to illustrate spatial variation in
% stellar pops and crowding conditions
\begin{figure}
\plotone{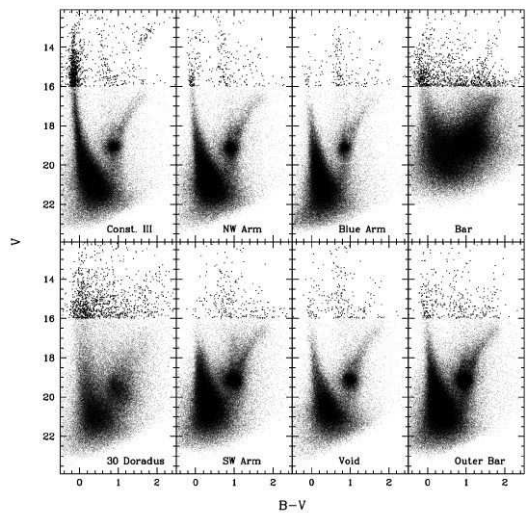}
\caption{$B-V$ color-magnitude diagrams for the same eight LMC
  subregions as shown in Figure~\ref{fig:extdist}.  Stars brighter
  than $V$=16~mag are shown with a larger point, to reveal the sparse
  distribution of these objects.  The dramatic spatial variation in
  the stellar populations evident in this plot motivates a
  comprehensive, spatially-resolved analysis of the LMC's
  SFH. \label{fig:cmdsample} }
\end{figure}              

\subsection{Deprojecting the LMC Disk}\label{sec:deproject}

\cite{vdm01a} showed convincingly that the stellar populations in the
LMC are distributed in a disk whose plane is inclined to
the plane of the sky by 34.7$^\circ$.  Over the region covered by the
MCPS, this inclination produces distance-modulus variations in the
photometry larger than $\pm0.1$~mag.  Prior to performing our StarFISH
analysis, we deproject the measured photometry to a common distance
modulus of 18.50~mag, using the geometry of the LMC disk reported by
\citeauthor{vdm02}.  We first transform each star's right ascension
and declination coordinates to a ($R$, $\Theta$) polar coordinate
system centered on the LMC (5.483333$^h$, -69.5$^\circ$).  Then the
line-of-sight distance to a star in the LMC plane, $D$ in kpc, is given by:

$$D = 50 + 0.873  R \tan(i)  \sin(\Theta - \Theta_0)$$

\noindent where $R$ is the angular separation of the star from the LMC
center (in degrees), $\Theta$ is the star's azimuthal angle, and $i$
and $\Theta_0$ are the inclination and position angle of the LMC disk,
as reported by \citeauthor{vdm01a}.  The value 0.873 represents the
linear separation (in kpc) corresponding to an angular separation of
one degree at a distance of 50~kpc.  The photometric correction
applied to the star is simply the difference between its in-disk
distance modulus and the canonical LMC distance modulus of 18.50~mag.

\subsection{Partitioning the Data}\label{sec:grid}

In Figure~\ref{fig:cmdsample}, we show $B-V$ CMDs for several
subregions from our MCPS catalog.  The stellar populations in these
regions show clear differences that are rooted in variations of the
SFH within the LMC.  For example, the region labeled ``Const. III''
has a very prominent upper main sequence and a rich population of
supergiants, indicating that this region has experienced a relatively
large star formation rate over the past several tens of millions of
years.  The photometric error properties also vary
substantially among these regions, driven primarily by variations in
the stellar surface density.  
Variations in the stellar content and photometric error 
distributions motivate the spatial partitioning of
our LMC catalog. The definition and characteristics of different regions
are treated in more detail in \S5.

% stellar density map, with hierarchical grid
\begin{figure}[htbp]
\plotone{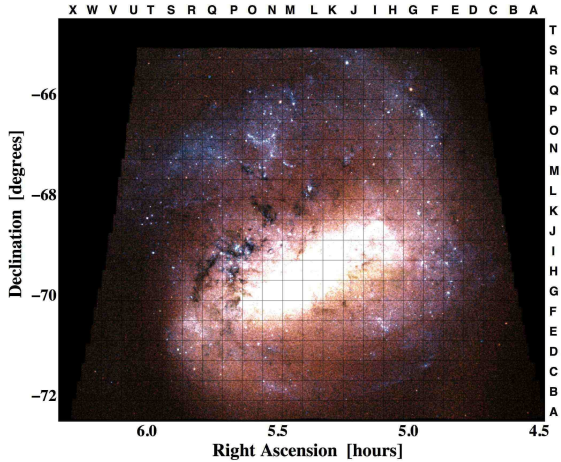}
\caption{A stellar flux density map derived from our $BVI$ photometry
  of the LMC.  Overplotted on this map, we show our gridding strategy
  for reconstructing the spatially-resolved SFH.  We divide the MCPS
  region into $24^\prime\times24^\prime$ cells.  Each of these cells
  may be further subdivided into a 2x2 grid of subregions, if the
  number of stars in the cell exceeds a threshold value.  In
  practice, most of the top-level grid cells are subdivided in this
  way. \label{fig:datagrid} }
\end{figure}           

We divide the MCPS photometry using a grid 
and perform an independent StarFISH analysis on each region.  This 
approach allows us to construct a spatially-resolved
map of the LMC's SFH.  The geometry of the grid (in particular, the size of the grid cells) is driven by several considerations.  To maximize our spatial resolution, 
we need to use the smallest possible regions. However, 
experience has taught us that StarFISH requires at least several thousand stars to avoid having
acceptable $\chi^2$ minima that lie beyond the error bounds of 
each minima. This number does in detail depend on the stellar populations
of the region, but the adopted number is generally applicable across our
MCPS. We must also be mindful to control the spatial
variations in the crowding interstellar extinction within each region.

Our gridding strategy is shown in Figure~\ref{fig:datagrid}.  The grid
has a two-level hierarchy: we first divide the catalog into a uniform
grid of 25$\times$20 cells, each spanning approximately 24$^\prime$ in
Right Ascension, and 24$^\prime$ in Declination.  If any of these
regions contain more than 25000 stars, then it is further subdivided
into a 2x2 sub-grid.  The final grid consists of 1380 subregions. We
calculate an independent SFH solution for each of these regions, using a
photometric error model drawn from one of our 17 ASTs that most
closely matches the target region in stellar surface density.  We also
ensure that the extinction properties of the target region's stars are
well matched by the synthetic CMD library by comparing the hot and
cool star extinction distributions as shown in
Figure~\ref{fig:extdist}.  If the extinction properties do not match,
we generate a new synthetic CMD library using the region's own
measured extinction properties.

% Gridding strategy for the CMDs for StarFISH analysis
\begin{figure}[htbp]
\epsscale{0.5}
\plotone{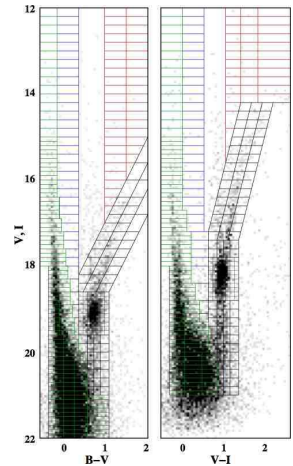}
\caption{Gridding strategy for the $B-V$ and $V-I$ CMD planes, for our
  StarFISH analysis.  The grid emphasizes the luminosity function of
  the main sequence (green boxes), the red giant branch and red clump
  (black boxes), and the red and blue supergiant sequences (red and
  blue boxes, respectively). \label{fig:cmdgrid} }
\end{figure}           

\subsection{Finding the Best-Fit Star Formation History Model}\label{sec:chi}

StarFISH employs a downhill simplex (``amoeba'') algorithm to
determine the best match between an observed CMD and a composite model
CMD, formed from the linear combination of all CMDs in the synthetic
library.  Again, these synthetic CMDs each represent the predicted
photometry for stars of a particular age and metallicity, given
observational parameters (distance modulus, photometric errors and
interstellar extinction properties) that are well-matched to the data.
Through linear combination, we form a composite model that can
represent any arbitrary mixture of ages and metallicities.  The
amplitude associated with each synthetic CMD in the composite model
describes the SFH of the population (\ie, its SFR and
metallicity distribution as a function of time).  These
amplitudes form the multidimensional parameter space in which we
search for the best fit SFH for the observed stellar population.  In
the present analysis, we have synthetic CMD pairs covering 9 age bins
and 4 metallicity bins, plus the 4 additional ``young'' age bins with
no metallicity variation, for a total of 40 dimensions in the
parameter space of the fitter.

The goodness-of-fit of any composite model CMD is evaluated by
statistical comparison to the observed CMD.  StarFISH can use a
standard $\chi^2$ fitting statistic, or the Lorentzian \citep{ols99} or
Poisson \citep{dol02} variants.  We have found very little practical
effect of this choice on the final best-fit determination, especially
if the CMDs are well populated with stars (so that the inherent
Poisson uncertainties of the CMD subregions approach the Gaussian
shape assumed by the $\chi^2$ statistic).  For the present analysis,
we choose the Poisson statistic.  Whatever statistic is used, the fit
determination involves dividing the CMD plane into small regions, and
counting the number of stars present in each region in both the
observed CMD and the composite model.  The fitting statistic is
minimized when the numbers of stars in the model's CMD regions match
those in the observed CMD regions most closely.  For the present
analysis, we use the CMD gridding strategy illustrated in
Figure~\ref{fig:cmdgrid}.  This gridding strategy emphasizes features in
the CMD pair that are sensitive to the SFH (such as the main sequence
luminosity function, the red giant branch, and supergiant sequences),
while ignoring regions that are likely dominated by contaminant
populations.

The amoeba algorithm is initialized at a random position in the SFH
parameter space, and it evaluates the fitting statistic for the
composite model CMD at that location.  It then takes a small step
along each parameter space dimension (\ie, varying each amplitude by a
small amount while holding the others fixed) to determine the local
gradient of the fitting statistic.  It takes a step in the
``downhill'' direction, and then recomputes the local gradient.  This
procedure iterates until a minimum is found.  As the amoeba descends
through the parameter space, the deviations taken to determine the
local gradient get smaller and smaller.  When a minimum is found, the
amoeba is reset to another random position within the SFH space, to ensure that it still
returns to the same minimum position.  In addition, we have added a
feature in which ``off-axis'' directions are probed for further
downhill gradient opportunities.  The stock amoeba algorithm only
searches along each parameter dimension when evaluating the gradient;
thus it can wrongly conclude that a minimum has been reached if
multiple amplitudes need to be simultaneously varied to achieve a
better fit. 

The SFH is often expressed as the SFR as a function of time.  However,
StarFISH does not directly determine the SFRs; fundamentally, it
determines the number of stars {\it present in the CMD} as a function
of age and metallicity.  In order to translate this to the SFR, we
need to perform a number of renormalizations.  First, we need to
convert from the number of stars that are observed in the CMD, to the
total number of stars present in the observed region.  This is
essentially a completeness correction, accounting for those stars that
are too faint to be observed.  This correction is a strong function of
age, and depends on an assumed IMF.  Once we have the total number of
stars present, we need an additional correction to get the total
number of stars that were originally formed.  This is a correction for
the fraction of the stellar population that has evolved away to
non-luminous end-states, and it is also a strong function of age.  For
example, we can observe nearly the entire stellar population if it is
a few Myr old, but if it is 12~Gyr old, then a significant fraction of
the stellar population no longer exists, and we must account for this
if we want to infer the number of stars formed 12~Gyr ago, from the
number of 12~Gyr stars that remain.  Once we have the total number of
stars formed at each age, we multiply by the average stellar mass
(given an IMF), and divide by the time interval covered by the age bin
to get the SFR in $M_\odot/yr$.

A more detailed discussion of the uncertainties in the method, tests of
the SFH reconstruction with known populations, comparison of observed
and synthetic CMDs, and the dependence of $\chi^2$ on undetermined parameters
such as binary fraction are presented by \cite{hz01,hz04}. Unless otherwise
specified, we have adopted the same parameters as adopted in \cite{hz04}.

\subsection{Estimating Fit Uncertainties}\label{sec:fiterrs}

Once the minimum of the fitting statistic has been determined,
StarFISH evaluates the fit uncertainties by exploring the parameter
space in the vicinity of this best-fit location.  Some SFH methods
simply rely on the statistics of the points evaluated on the way to
reaching the minimum to estimate the parameter uncertainties.
However, we find that a systematic exploration of the parameter space
surrounding the best-fit location does a much better job of
determining realistic uncertainties.  With 40 parameters, it is not
possible to completely explore the parameter space, but we have
developed a multi-step procedure that allows us to examine the
independent variance of each amplitude, the covariance of pairs of
amplitudes, and the general covariance of all amplitudes.  In all
cases, we step away from the best-fit parameter space location along a
single parameter-space direction until the fit worsens to the point
that we have reached the 68\% (1 $\sigma$) confidence interval.  Along
the way, we keep track of the maximum deviation of each amplitude from
its best-fit value that remains within the 68\% confidence interval.
In the first phase of the procedure, we determine the independent
variance of each amplitude by both increasing and decreasing each
amplitude while holding all others fixed, until the 68\% confidence
limit is reached.  In the second phase, we explore the pairwise
covariance of adjacent amplitudes.  For each pair of adjacent
amplitudes, we allow them to vary against each other while all other
amplitudes are held fixed, until the 68\% confidence limit is reached.
This procedure is done both in age and metallicity.
In the third phase, we explore general covariance by choosing a
completely random parameter space direction and stepping away from the
best-fit location until the 68\% confidence limit is reached.  The
random-direction search is repeated 30,000 times in an attempt to
adequately sample the topology of the local parameter space.
Formally, 30,000 iterations is grossly inadequate to cover a
40-dimensional parameter space.  However, we find that the local
topology tends to be sufficiently smooth that exploring the
independent variance of each amplitude, as well as the pairwise
variance of all pairs of amplitudes, is usually enough to provide a
reasonable picture of the fit uncertainties.  In other words, it is
rare that our phase 3 explorations result in a further increase of the
confidence intervals over and above what phases 1 and 2 determined.

% Map of subregions given special attention
\begin{figure}[htbp]
\epsscale{1.0}
\plotone{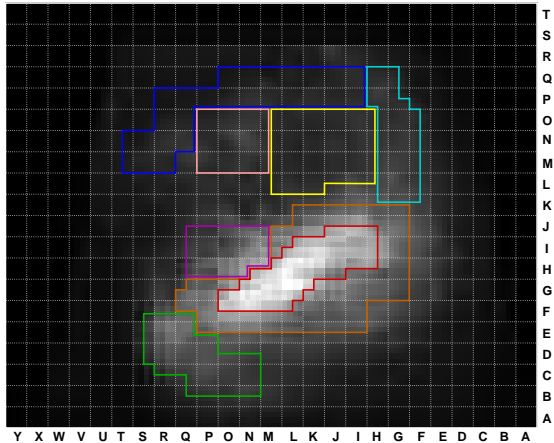}
\caption{A star-count map of the LMC, in which the pixel value in each
  grid region is proportional to the number of stars present in that
  region.  The map highlights several collections of regions for which
  we will examine the SFH in detail: the LMC Bar (red), the Outer Bar
  (orange), 30~Doradus (magenta), the ``Southeast Arm'' (green), the
  ``Northwest Arm'' (cyan), the ``Blue Arm'' (blue), Constellation~III
  (pink), and the ``Northwest Void'' (yellow).
\label{fig:regmap} }
\end{figure}

\subsection{Using HST Results to Constrain the Bar's Ancient SFH}\label{sec:hstanchor}

Our MCPS photometry is relatively shallow, and barely reaches the
ancient MSTO which occurs around $V=21$~mag for stars in the LMC.  In
the crowded bar regions, our photometry is significantly shallower,
perhaps a magnitude brighter than the old MSTO.  Because most of the
age information for old stellar populations is provided by the MSTO,
it is difficult to extract the early SFH of the LMC from our MCPS
photometry, especially for the bar region.  For this reason, we
restrict StarFISH to fitting a single age bin covering all ages older
than 4~Gyr in the bar region.

\begin{figure}[htbp]
\plotone{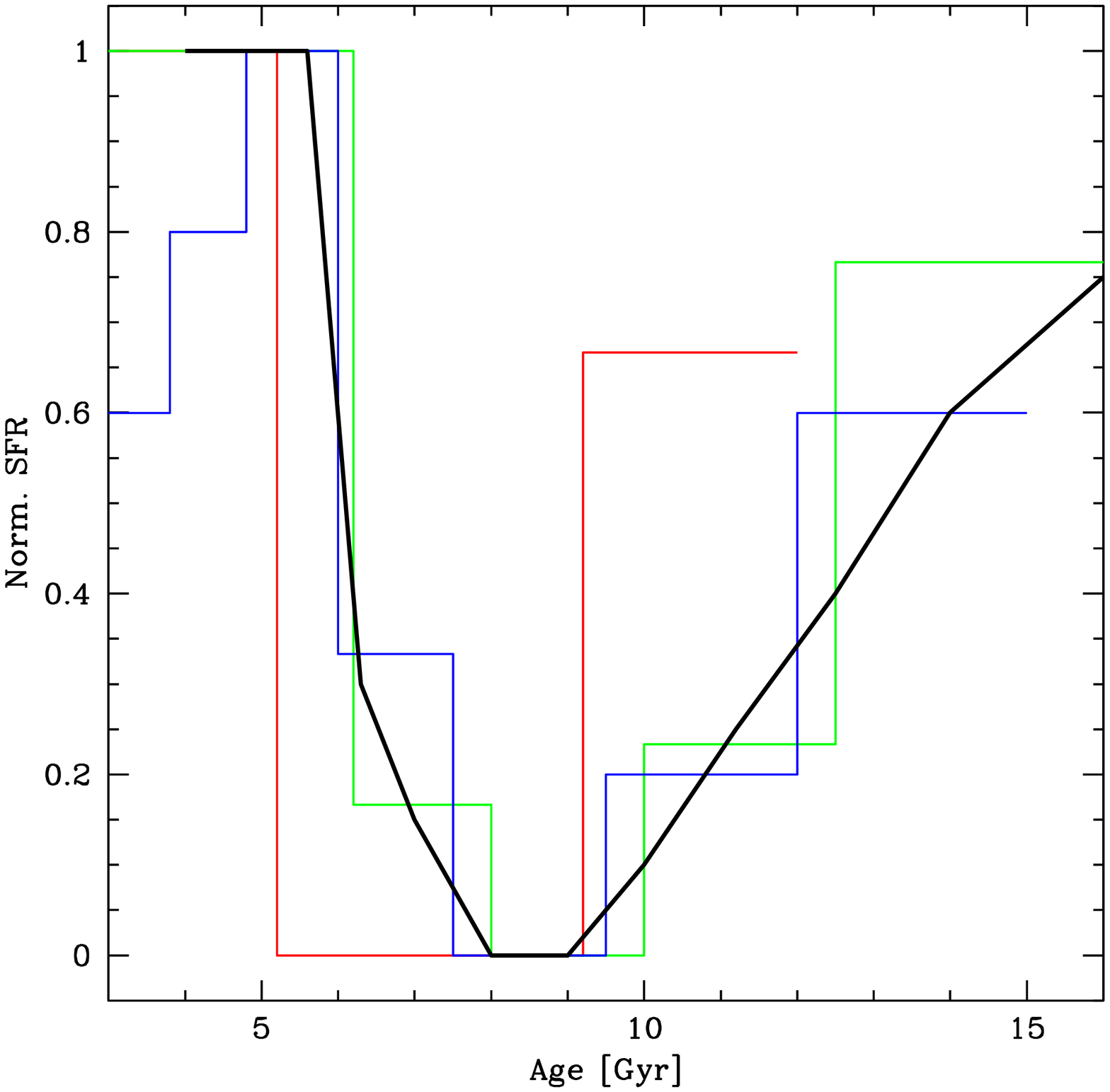}
\caption{The SFH solutions for ages $>4$~Gyr from three independent
  studies of the LMC Bar, using HST imaging.  The solutions are:
  \cite{ols99} (red line), \cite{hol99} (green line), and \cite{sh02}
  (blue line).  The three solutions show remarkable agreement,
  especially when considering that \citeauthor{ols99} chose a
  relatively young age limit of 10~Gyr.  In the present analysis, we
  will require that our SFH solutions conform to the black curve for
  ages older than 4~Gyr.  In this way, we rely on the deep HST imaging
  to supplement our relatively shallow ground-based
  data. \label{fig:hstsfh} }
\end{figure}

Within the bar (see Figure \ref{fig:regmap}), we use the 
information for old ages from published SFH
solutions based on deep HST data, as mentioned in
Section~\ref{sec:intro}.  The HST fields, while tiny, should sample
the old SFH fairly because orbital mixing will have erased any spatial
variation.  The derived SFH in the three published studies that used
deep HST imaging of LMC bar fields 
\citep{ols99, hol99, sh02} agree
remarkably well (see Figure~\ref{fig:hstsfh}).  Each finds an initial
epoch of star formation that is extinguished by 10~Gyr ago, and then
resumes about 5~Gyr ago.  The shape of the \citeauthor{ols99} solution
is somewhat different than the other two, but this disagreement is
likely due to the lower time resolution and younger age limit used in
that study.  We adopt a consensus SFH based on these HST results to
represent the portion of our SFH solutions older than 4~Gyr.  In other
words, we use StarFISH to determine the total number of stars older
than 4~Gyr in a given LMC bar subregion, but those stars are distributed
in age according to the consensus HST SFH solution.

% SFH results for HST studies of Bar fields

To check the validity of our HST-anchored solutions, we performed an
alternative set of solutions for all 1380 subregions in the LMC map,
in which we did not use the HST results to constrain the old SFH.  In
this case, we simply extend our $\Delta \log (t)$=0.2 age-binning
strategy to include bins with $\log (t)=$9.8, 10.0 and 10.2.  We find
that this alternative set of solutions produces obvious
spatially-dependent artifacts in the early panels of the SFH map.
However, for ages $\le 4$~Gyr, the SFH map is indistinguishable from
that of the HST-anchored solutions.  Because of the artifacts present
in the early bins, we will restrict ourselves to the set of solutions
constrained to match the HST results for ages older than 4~Gyr for the
remainder of the paper, unless otherwise noted.

\section{Results: A Map of the LMC's Star Formation History}\label{sec:sfhmap}

\begin{figure*}[htbp]
\epsscale{0.85}
\plotone{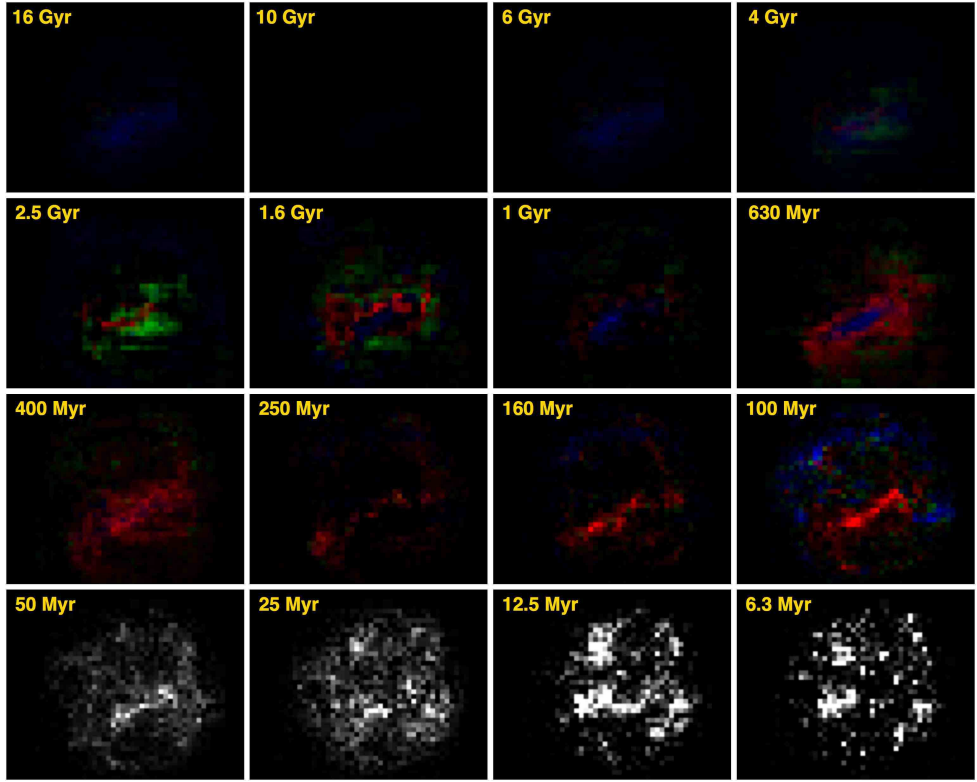}
\caption{The map of the LMC's SFH, based on the StarFISH solutions for
  over 1300 subregions.  The pixel value for each region is
  proportional to the star formation rate at that location, color
  coded for the metallicity (with Z=0.001 shown as blue, Z=0.004 shown
  as green, and Z=0.008 shown as red).  There is no color-coding for
  log(Age)$<8.0$, because the fitter is not given multiple metallicity
  bins for these young ages.\label{fig:sfhmap} }
\end{figure*}

\begin{figure*}[htbp]
\epsscale{0.85} \plotone{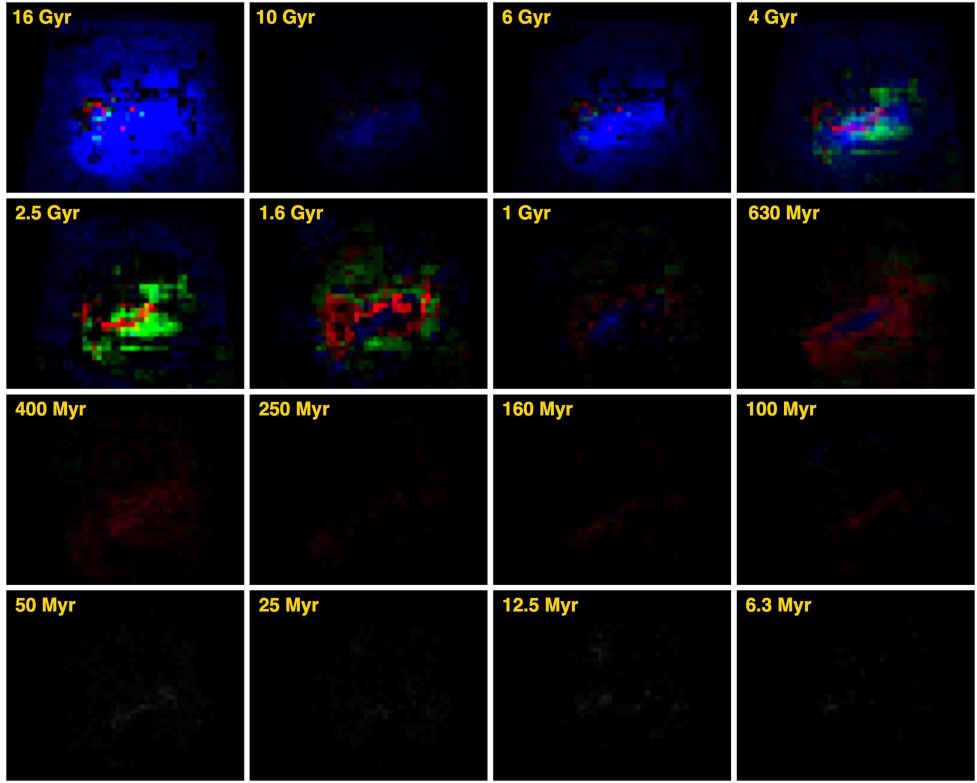}
\caption{The map of the LMC's SFH, based on the StarFISH solutions for
  over 1300 subregions. In this version of the map, the pixel values
  are proportional to the total stellar mass formed, rather than the
  star formation rate.\label{fig:sfhmap_stellarmass} }
\end{figure*}

We have performed SFH solutions 
for nearly 1400 LMC subregions.  Putting
all of these solutions together allows us to construct a 
map of the total SFH of the LMC, which is shown in
Figure~\ref{fig:sfhmap}.  The map shows the entire MCPS survey region
covering the central $8^\circ\times8^\circ$ of the LMC.  Each panel
shows the star formation rate for a particular time-step in the LMC's
history.  The pixel values are proportional to the SFR at that time,
color-coded for the metallicity.  Because the time interval covered by
each panel increases logarithmically with time, displaying the star
formation rate does not provide an intuitive grasp of the age at which
most stars were formed.  We therefore also provide an alternative view
of the SFH map, in which the pixel values are proportional to the
total stellar mass formed in each timestep
(Figure~\ref{fig:sfhmap_stellarmass}).  In this view, it is apparent
that a significant portion of the LMC's stellar mass had been
assembled by 10~Gyr ago.

\begin{deluxetable*}{rrrrrrrrrrrrr}
\tabletypesize{\scriptsize}
\tablecolumns{10}
\tablewidth{0pt}

\tablecaption{The Star Formation History of the LMC \label{tab:smcsfh}}
\tablehead{
    \colhead{} & \multicolumn{3}{c}{Z = 0.008} & \multicolumn{3}{c}{Z = 0.004} &
        \multicolumn{3}{c}{Z = 0.0025} & \multicolumn{3}{c}{Z = 0.001} \\
        \\
        \cline{2-4} \cline{5-7} \cline{8-10} \cline{11-13}  \\
    \colhead{log(age)} &
        \colhead{$SFR$} & \colhead{$SFR_{l}$} & \colhead{$SFR_{u}$} &
        \colhead{$SFR$} & \colhead{$SFR_{l}$} & \colhead{$SFR_{u}$} &
        \colhead{$SFR$} & \colhead{$SFR_{l}$} & \colhead{$SFR_{u}$} &
        \colhead{$SFR$} & \colhead{$SFR_{l}$} & \colhead{$SFR_{u}$}
}
\startdata
\multicolumn{13}{c}{Region AA  ( 04$^h$ 31$^m$,   $-$72\arcdeg\ 17\arcmin )} \\
\cline{1-13} \\
10.20&0&0&5&0&0&7&0&0&12&39&28&53\\
10.00&0&0&4&0&0&8&0&0&13&24&13&37\\
9.80&0&0&5&0&0&8&0&0&14&17&5&31\\
9.60&0&0&6&0&0&11&0&0&15&0&0&12\\
9.40&0&0&12&67&50&87&191&174&210&14&0&28\\
9.20&0&0&16&0&0&12&0&0&13&0&0&11\\
9.00&4&0&17&0&0&10&0&0&8&0&0&8\\
8.80&7&0&21&0&0&11&0&0&13&17&7&30\\
8.60&0&0&12&11&0&28&0&0&16&0&0&11\\
8.40&0&0&9&0&0&9&0&0&10&0&0&11\\
8.20&0&0&9&0&0&14&0&0&18&10&0&32\\
8.00&0&0&12&0&0&19&9&0&37&5&0&30\\
7.70&8&0&27&0&0&0&0&0&0&0&0&0\\
7.40&12&0&38&0&0&0&0&0&0&0&0&0\\
7.10&0&0&28&0&0&0&0&0&0&0&0&0\\
6.80&0&0&46&0&0&0&0&0&0&0&0&0\\
\enddata
\tablecomments{The complete version of this table is in the
electronic edition of the Journal. The printed edition contains
only a sample. SFRs are in units of 10$^{-6} M_\odot {\rm yr}^{-1}$}
\end{deluxetable*}

% map of the LMC's SFH

Figures~\ref{fig:sfhmap} and \ref{fig:sfhmap_stellarmass} show that
the LMC's SFH has rich structure both spatially and temporally.  Due
to dynamical mixing and the longer time intervals covered, the older
time bins are generally spatially and temporally smoother than the
younger bins.  In the youngest bins, we can associate peaks in the map
with well-known star-forming regions in the LMC.  For example, the
concentration to the left of center and slightly below is 30~Doradus,
and the concentration directly above 30~Doradus is LMC
Constellation~III.

% map of the LMC's SFH, expressed as stellar mass formed per timestep

\begin{figure}[htbp]
\epsscale{0.85}
\plotone{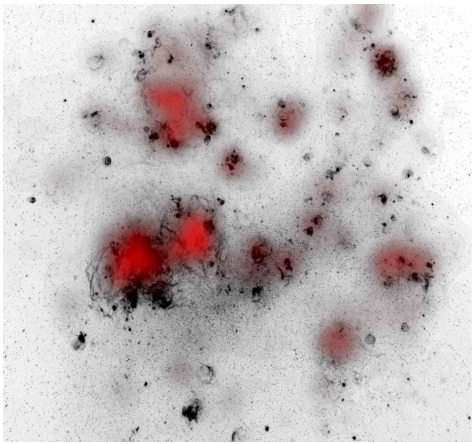}
\caption{The correlation of the recent (Age $<12.5 $~Myr) star-formation
  activity in the LMC based on our analysis, with the H$\alpha$ image
  of the LMC from the Magellanic Clouds Emission Line Survey (MCELS)
  \citep{poi08}. \label{fig:youngmap} }
\end{figure}

In Figure~\ref{fig:youngmap}, we show the spatial correlation of the
young (Age $\le$ 12.5~Myr) star-formation activity as derived from our
analysis and the H$\alpha$ emission in the LMC from the Magellanic
Clouds Emission Line Survey \citep{poi08}.  As expected, H$\alpha$
flux and recent star-formation activity correlate because both
quantities are causally linked to the presence of young stars.
However, the correlation is not perfect: there are regions that have
large H$\alpha$ flux, but show little or no signal in the recent SFH
map and there are hotspots in the recent SFH map that show little or
no H$\alpha$ flux.  Some of the non-correlation can be understood from
the fact that H$\alpha$ flux is not purely a tracer of star formation
activity; it requires {\em both} recent star formation {\em and} the
presence of significant amounts of hydrogen gas.  For example, the
large region of recent star-formation activity visible in the northern
disk in Figure~\ref{fig:youngmap} is the Constellation~III region.  We
know from tracers of the interstellar medium (such as H I and 8$\mu$m
emission) that Constellation~III is embedded in a cavity from which
the gas has been evacuated \citep{kim98, mei06}, making H$\alpha$
emission here impossible.  On the other hand, there are many regions
in Figure~\ref{fig:youngmap} in which we see a localized peak in the
H$\alpha$ flux with no corresponding peak in the recent SFH map.  To
understand these cases, we point to the different spatial resolutions
of the SFH map and the H$\alpha$ image.  A small cluster or
OB~association can lead to a bright HII region with only a few
photoionizing stars.  However, a small handful of young stars will not
be statistically significant enough to sway the SFH solution.

% correlation of young SF activity with Halpha
                                                                                                                           
The LMC bar has had a distinct SFH from that of the general LMC disk.
The bar dominates in some of the panels in Figure~\ref{fig:sfhmap}
(16~Gyr, 6~Gyr, 630~Myr, 400~Myr, 160~Myr), but is largely absent from
others (1.6~Gyr, 250~Myr, 25~Myr, 12.5~Myr, 6.3~Myr).  At 2.5~Gyr,
there is centrally-concentrated activity in the map, but unlike that
seen in the younger panels, its structure does not closely resemble
the current contours of the bar.

At 160 and 100~Myr, there is a ``Northern Arm'' feature composed of
low-metallicity star-formation activity in the outskirts of the
northern LMC disk.  The possibility of recent low-metallicity star
formation activity that is coherent across several kpc in the LMC disk
is intriguing.  However, metallicity variations become photometrically
degenerate for younger stellar populations, and in fact we chose to
eliminate metallicity variations for ages younger than 100~Myr in
order to avoid these degeneracies.  What features in the CMD are
driving the fitter to conclude that the northern outskirts of the LMC
disk saw low-metallicity star formation 100~Myr ago?  The likely
answer is that the mean color of the faint main sequence is slightly
bluer in these regions than in other regions.  Overplotting the
Z=0.001 and Z=0.008 isochrones on a CMD for an age of 100~Myr shows
that the upper main sequences are nearly coincident, but toward
fainter magnitudes, the metal-rich isochrone shows a deviation to the
red.  By $M_V$=2.5~mag ($V$=21~mag for the LMC), the deviation in
$(B-V)$ is almost 0.2~mag.  While the supergiant sequences in the
isochrones show substantial differences between Z=0.001 and Z=0.008,
these evolutionary stages are so poorly populated compared to the main
sequence that they do not carry much weight in determining the fit.
As a test, we selected two adjacent regions from the LMC SFH map that
show similar activity in the 100~Myr panel (with one dominated by
Z=0.001 stars, and the other by Z=0.008 stars).  Indeed, the faint
main sequence stars in the metal-poor region are 0.1~mag bluer, on
average, than the stars in the metal-rich region.  To the extent that
the faint-end color variations are considered real, metallicity
variations are one plausible, but not unique, explanation for them.
Spectroscopic follow-up is needed to confirm these apparent
metallicity variations among stars aged 100--160~Myr in our map.

\section{Discussion}\label{sec:discuss}

\subsection{The LMC's Total Star Formation History}\label{sec:totsfh}

\begin{figure}[htbp]
\epsscale{0.85}
\plotone{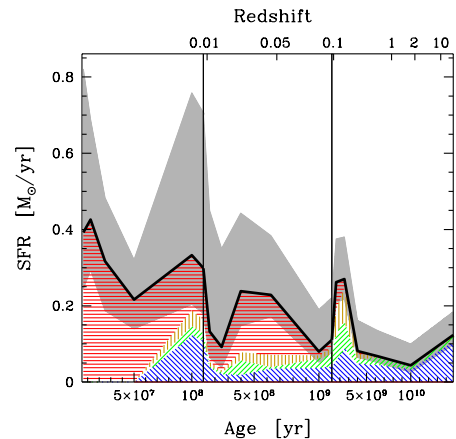}
\caption{The total SFH of the LMC, computed by summing over all 1376
  regions covering the MCPS survey region.
   The time axis is shown with a linear
  scale that is broken into three segments: the left panel covers
  $\sim$100~Myr, the middle panel covers $\sim$1~Gyr, and the right
  panel covers $\sim$14~Gyr.  The best-fit star formation rate (SFR)
  as a function of age is shown with a thick black line; the
  uncertainty on the fit (including covariance between age bins) is
  shown as a grey shaded envelope.  The distribution of metallicity at
  each age is shown by the mix of colors below the SFR line (Z=0.001
  in blue and downward sloping ; Z=0.0025 in green and upward sloping;
   Z=0.004 in orange and vertical ; Z=0.008 in
 red and horizontal). \label{fig:sfh_lmc} }
\end{figure}

Combining the SFH solutions from all 1380 LMC regions, we construct
the SFH for the entire LMC (Figure~\ref{fig:sfh_lmc}).  In the Figure
we also show an alternative representation of the SFH: the cumulative
stellar mass formed in the LMC as a function of time, which shows that
half of the LMC's stellar mass was assembled 5~Gyr ago.  Note that the
stellar mass in this plot includes the initial mass of all stars
formed, including stars that have already evolved to non-luminous
end-states and returned large fractions of their mass to the
interstellar medium.

% total SFH for the LMC

One of the most dramatic features in the global SFH solution is the
quiescent epoch spanning roughly 5--12~Gyr ago.  The presence of this
quiescent epoch is imposed by our decision to tie the old SFH to the
HST results (see Figure~\ref{fig:hstsfh}).  However, the quiescent
epoch is also present in the total SFH derived from the alternate set
of solutions that were {\it not} tied to the HST results, so it is not an artifact of the
adopted HST solutions, but a robust feature of the LMC's history,
reflected in both its cluster and field populations.  After an initial
epoch of star formation, the LMC all but ceased forming stars,
sustaining an average star formation rate of only 0.02~$M_\odot
yr^{-1}$ over the interval between 12 and 5~Gyr.  Following the
quiescent epoch, star formation resumed throughout the galaxy, and it
has remained active ever since, with an average rate of 0.2~$M_\odot
yr^{-1}$.  It is tempting to speculate that something dramatic
happened to the LMC 5~Gyr ago to precipitate the global resumption of
star formation, such as a merger with a gas-rich dwarf galaxy, or a
gravitational interaction with the Milky Way.

The total SFH solution in Figure~\ref{fig:sfh_lmc} shows episodes of
enhanced star-formation activity at 12~Myr, 100~Myr, 500~Myr, and
2~Gyr ago.  These episodes represent deviations from the long-term
average star formation rate of up to a factor of two.  However, as our
age resolution becomes coarser with increasing age, ever larger
deviations in the star formation rate can remain undetected if they
occurred on a sufficiently short timescale.

% mass build up
\begin{figure}[htbp]
\epsscale{0.85}
\plotone{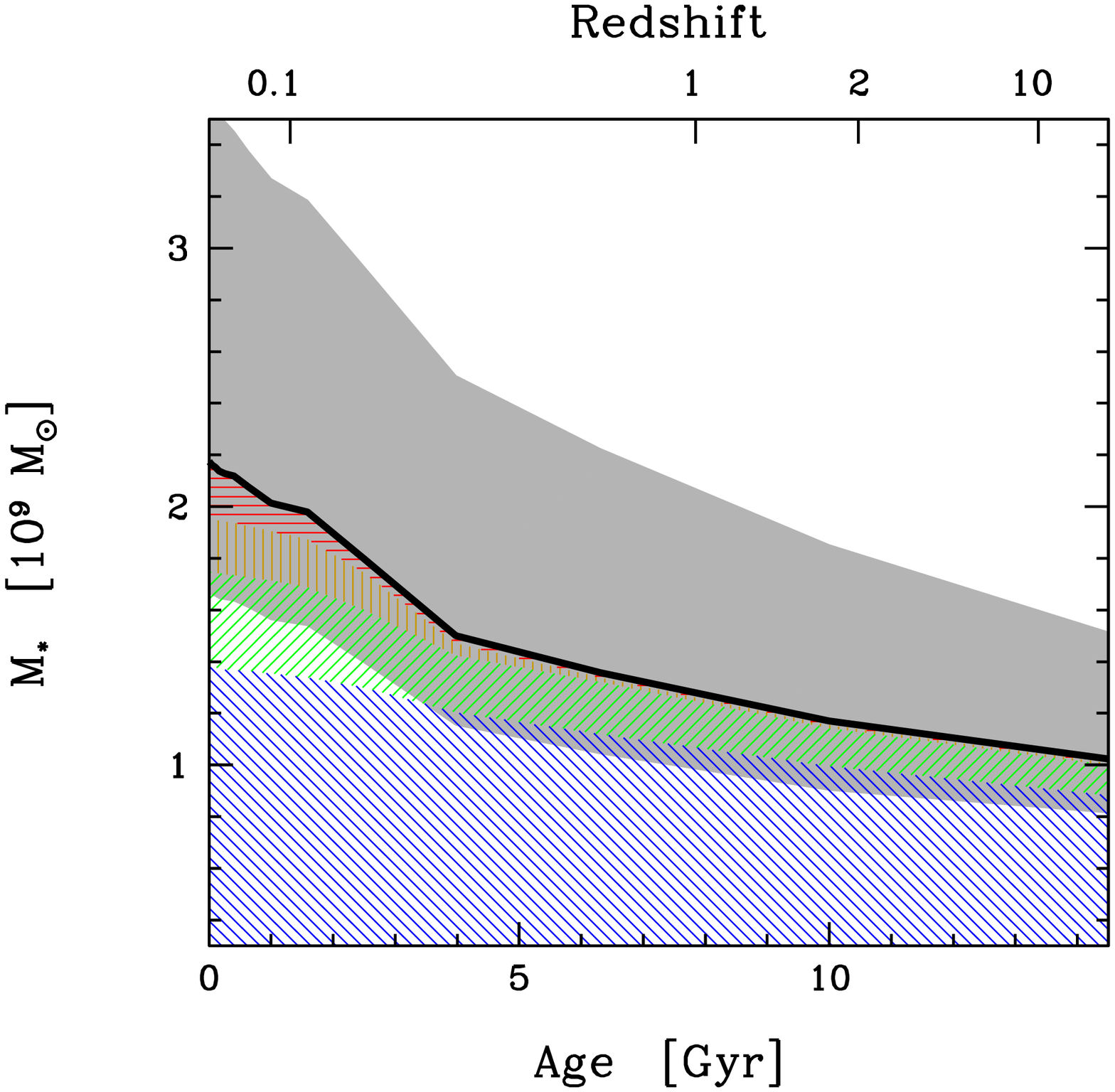}
\caption{The cumulative stellar mass formed in the the LMC. Shading and colors
same as in previous Figures although the time axis is now linear.
 \label{fig:sfh_cumm} }
\end{figure}
                                                                                                                           
As noted by many for at least 20 years \citep[cf.][]{vdb81,nw2000}, the peaks in the SFR did not occur uniformly throughout the LMC
(Figure~\ref{fig:sfhmap}) and our study enables a detailed temporal and
morphological study of this behavior.  For example, the activity at 100~Myr is
concentrated along the present-day bar, while the activity at 2~Gyr is
centrally concentrated and does not follow the contours of the
present-day bar.  We examine the SFH solutions of eight localized
features in the LMC, some of which are well-known structures (such as
the bar), and others which are only distinguishable in our SFH map.
The eight regions are outlined in Figure~\ref{fig:regmap}.

\subsubsection{The LMC Bar}\label{sec:bar}

The LMC bar is an enigmatic cigar-shaped structure that dominates
optical images of the LMC.  This feature is often thought of as a
massive dynamical bar, which are common at the centers of spiral
galaxies.  However, for a number of reasons, it is difficult to
understand the LMC bar as a dynamical structure embedded in the LMC
disk.  Most dramatically, the bar is absent in images that trace the
interstellar medium of the LMC, such as HI \citep{kim98}, H$\alpha$
\citep{poi08}, CO \citep{miz01}, or 8$\mu$m and 24$\mu$m emission
\citep{mei06}.  A dynamical bar is the manifestation of a
gravitational instability in a galaxy disk, and should therefore sweep
all types of matter -- stars, gas and dust.  Another, perhaps related,
source of ambiguity is that present-day star formation is not
concentrated in the bar, as often happens in barred spiral galaxies.
Dynamical bar structures funnel gas and dust into the center of the
galaxy \citep{mac04}, leading to a nuclear starburst \citep[\eg
][]{sch06}.  In the case of the LMC, the most active star-forming
regions in the LMC (including 30~Doradus) lie well outside the bar.
Lastly, differential distance measurements to the bar and disk suggest
that the bar is closer to us by as much as 0.5~kpc \citep{nik04}.
The complicated interaction history, both with the SMC and our Galaxy, 
provide many opportunities for complicated dynamical effects and
distorted morphological features \citep[cf.][]{ss03}.

% SFH of the LMC Bar
\begin{figure*}[htbp]
\epsfig{figure=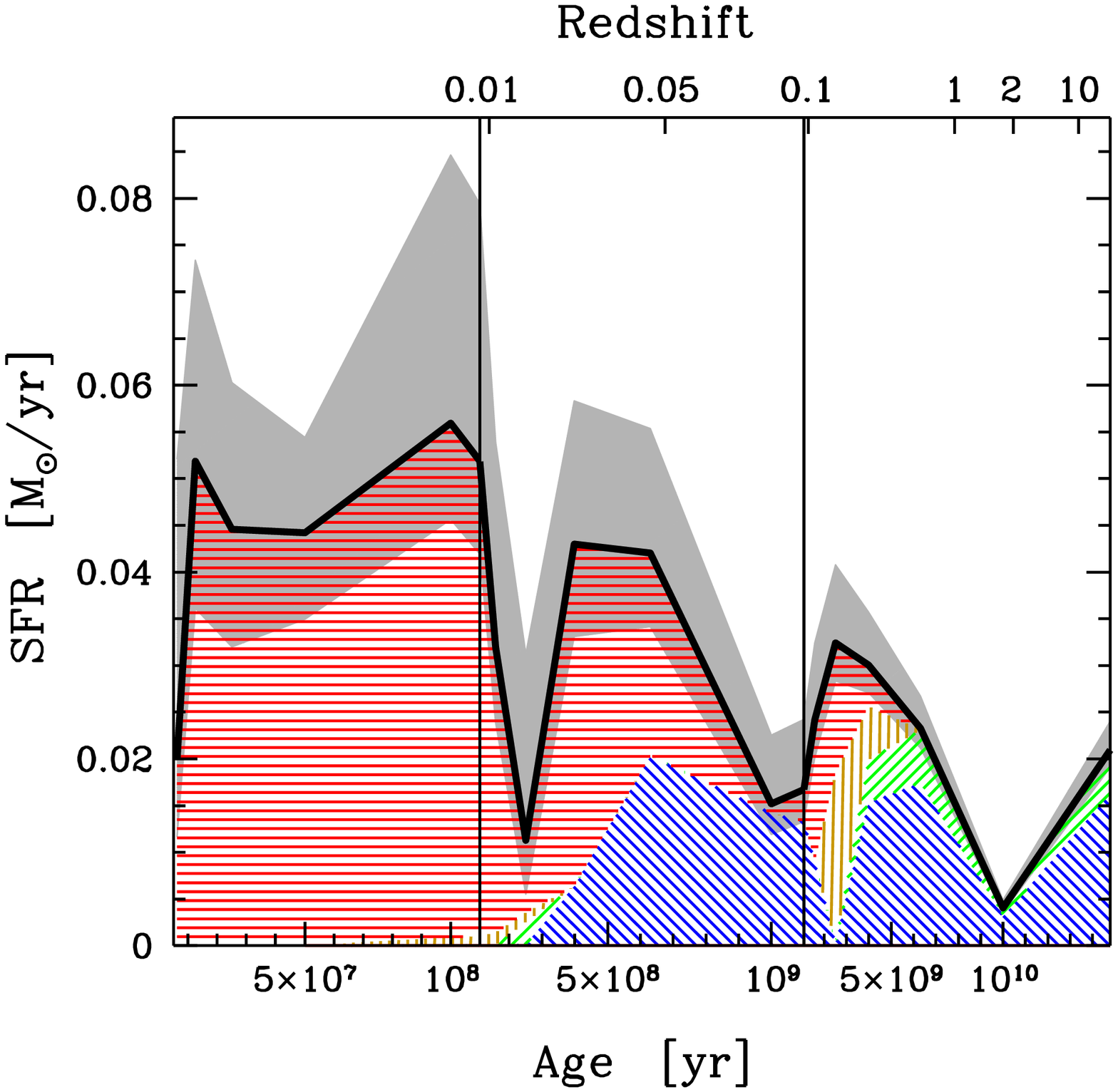,width=2.9in,angle=0}
\epsfig{figure=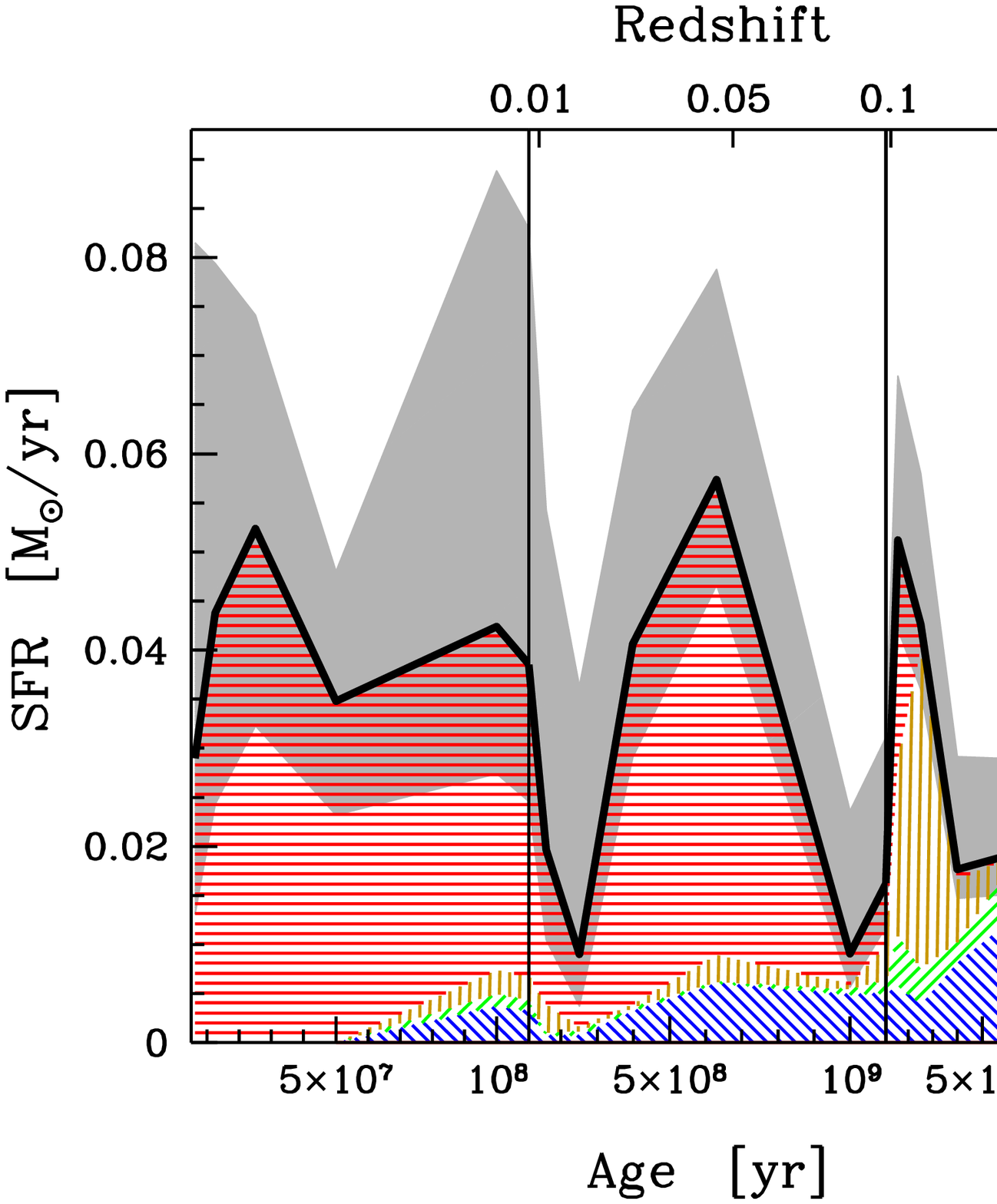,width=2.9in,angle=0}
\caption{The integrated SFH solutions for the LMC Bar (left) and Outer
  Bar (right) regions.  The shading and colors are the same as in
  Figure~\ref{fig:sfh_lmc}. \label{fig:sfh_bar} }
\end{figure*}

% SFH of LMC ``arms''
\begin{figure*}[htbp]
\epsfig{figure=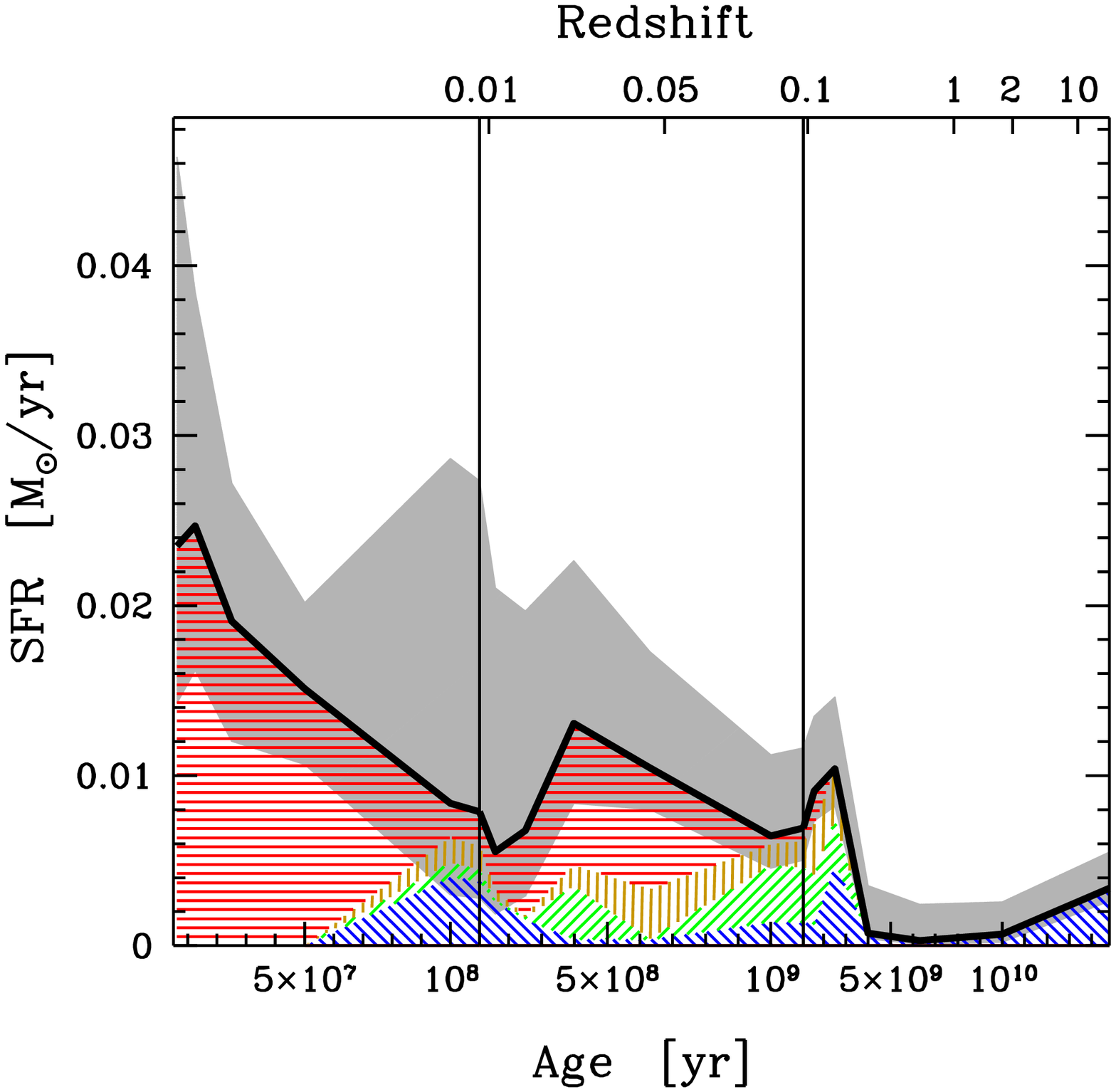,width=2.9in,angle=0}
\epsfig{figure=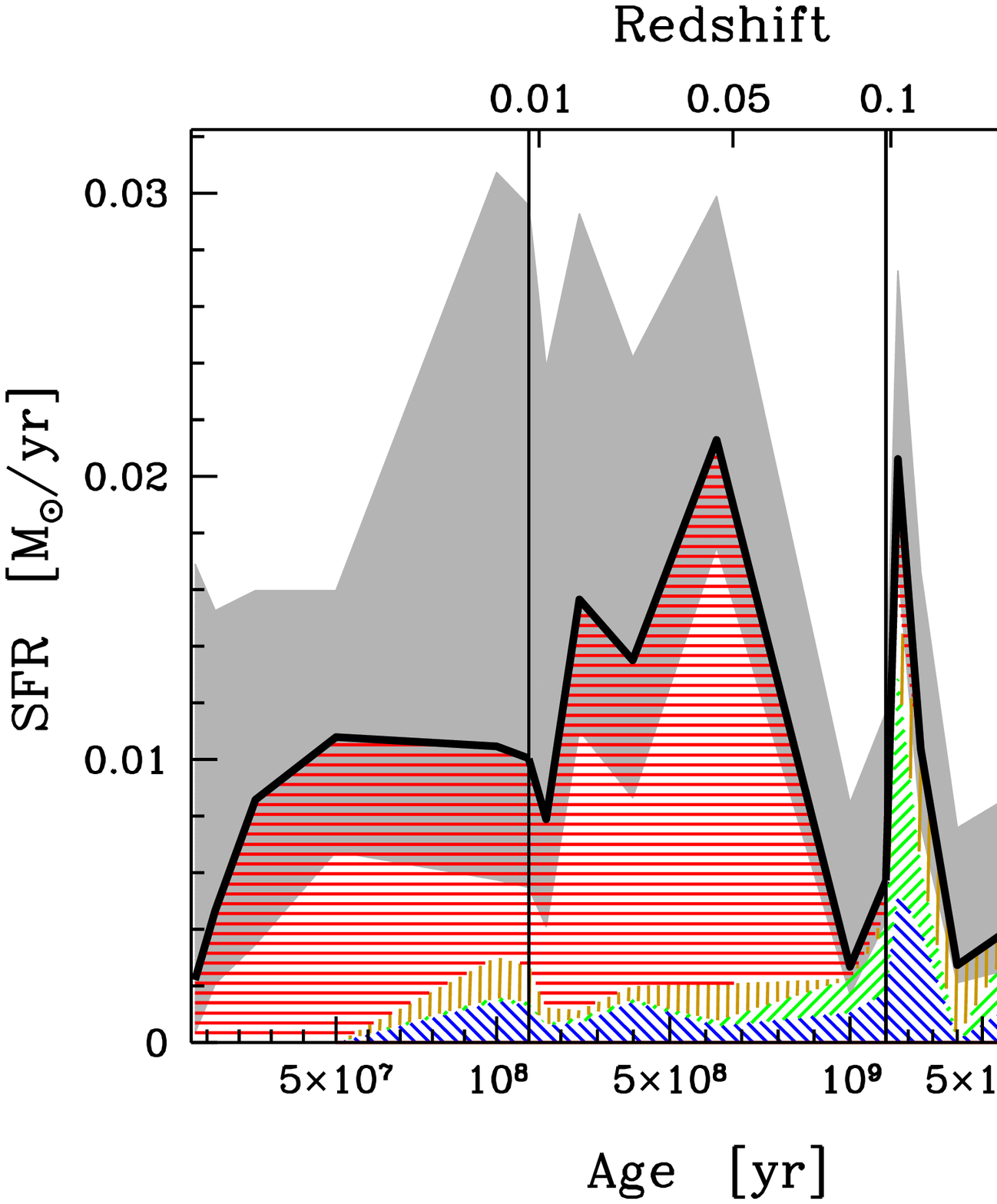,width=2.9in,angle=0}
\caption{The integrated SFH solutions for the Northwest Arm (left) and
  the Southeast Arm (right) regions.  The shading and colors are the
  same as in Figure~\ref{fig:sfh_lmc}. \label{fig:sfh_arms} }
\end{figure*}                    

\cite{zar04b} proposed that the LMC bar may actually be a bulge
population that is half-obscured by the optically-thick disk of the
LMC.  This hypothesis can explain many of the bar's properties, but
the required disk opacity is challenged by extinction measurements in
the LMC, by the distribution of stellar flux in infrared images,
which are far less obscured by dust, and by the stellar kinematics, which
are normal \citep{cole05}. 

These mysteries motivate us to look for answers in the bar's SFH.  In
the SFH map of the entire LMC (Figure~\ref{fig:sfhmap}), the bar shows
intermittent periods of activity and quiescence throughout the LMC's
history, as has been noted previously \citep{hardy, sh02}.  
It had particularly active episodes 5~Gyr, 500~Myr, and
100~Myr ago.  At other times (\eg, 2.5~Gyr and 12~Myr), there is
significant star-formation activity in the bar, but the distribution
of the star-formation activity at these ages does not match the
present-day structure of the bar.  At 2.5~Gyr, the
centrally-concentrated activity is oriented along a position angle
that is closer to the East-West direction than the present-day bar.
This offset may mean that the 2.5~Gyr old stellar population has
little to do with the bar.  Similarly, the activity at 12~Myr is
dominated by 30~Doradus and the Constellation~III region, which are
not related to the bar.  While there is corresponding activity near
the center of the bar region, it does not pervade the entire bar, as
it does at 100~Myr.

The integrated SFHs of the inner and outer bar regions are shown in
Figure~\ref{fig:sfh_bar}, although given the questions about the nature
of the bar defining the bar population is difficult \citep[cf.][]{cole05}.  
There is striking coherence in the timing
of star-formation events among the bar, arm
(Figure~\ref{fig:sfh_arms}), and void (Figure~\ref{fig:sfh_void})
regions.  Whatever its nature, the bar has been an integral part of
the LMC for most of its history.  The biggest distinguishing features
of the bar's SFH are that its quiescent periods at 250~Myr and 1~Gyr
are relatively deeper than in other parts of the LMC and the peak in
the SFR at 5~Gyr is relatively stronger. 

\subsubsection{The LMC Arms}\label{sec:arms}

While the LMC is clearly a disk galaxy, the evidence for spiral arms
in the LMC that are the result of a spiral density wave is tenuous at
best.  In this Section, we refer to the LMC's ``arms'' as the
structures extending off the ends of the LMC bar that are
morphologically similar to spiral arms (see Figure~\ref{fig:regmap}),
without implying that there is necessarily a spiral density wave in
the LMC's disk.

The integrated SFHs of the northwest and southeast LMC arm regions
(Figure~\ref{fig:sfh_arms}) are remarkably different.  The northwest
arm region had very little star-formation activity prior to 5~Gyr ago,
and it also lacks the quiescent period at 1~Gyr seen in most other
regions.  The activity at 100~Myr seen in most other regions also
seems to be suppressed here, and its recent SFH is characterized by a
steady increase in the SFR since 100~Myr ago.  In the southeast arm,
we see one of the strongest signatures of the short-term burst of
activity at 2.5~Gyr.  Most remarkably, the recent star-formation
activity in the southeast arm peaked 600~Myr ago and has been
declining since.  These differences suggest that these structures are
probably not the sole result of a spiral density wave in the LMC disk
for which we would expect to see more symmetry.

\subsubsection{The ``Blue Arm''}\label{sec:bluearm}

We define the region we call the ``blue arm'' to enclose the coherent
low-metallicity star-formation activity in the LMC's northern disk
100~Myr ago (Figure~\ref{fig:sfhmap}).  In Section~\ref{sec:sfhmap},
we argued that the low metallicity of this feature should be treated
with some skepticism because metallicity variations produce only
subtle differences along the main sequence of younger stellar
populations.  While we still recommend this skepticism, it is
interesting to note that the integrated SFH of the blue arm region
shows that the metallicity here has been relatively low throughout its
history.

% SFH of the ``blue arm'' feature
\begin{figure}[htbp]
\epsscale{0.85}
\plotone{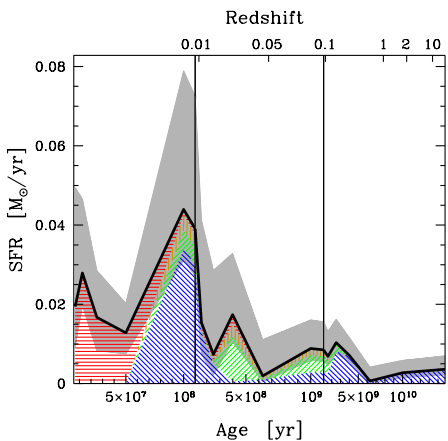}
\caption{The integrated SFH solution for the Blue Arm region.  The
  shading and colors are the same as in
  Figure~\ref{fig:sfh_lmc}. \label{fig:sfh_bluearm} }
\end{figure}
                                                                                                                           
Even if the low metallicity in this region is a photometric artifact,
the blue arm is still remarkable as the remains of a major star
formation event in the outer disk of the LMC 100--160~Myr ago.  The
total stellar mass formed during this event was roughly a factor of
five greater than the stellar mass formed in Constellation~III or
30~Doradus.  The blue arm is less spatially concentrated than these
younger structures, probably due to dynamical mixing during the
100~Myr since its formation.

\subsubsection{Constellation~III}\label{sec:consIII}

\begin{figure}[htbp]
\epsscale{0.85}
\plotone{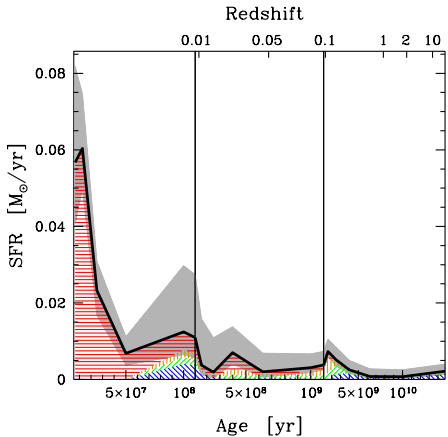}
\caption{The integrated SFH solution for the Constellation~III region.
  The shading and colors are the same as in
  Figure~\ref{fig:sfh_lmc}. \label{fig:sfh_consIII} }
\end{figure}

In \cite{hz08}, we performed a detailed analysis of the SFH of
Constellation~III, based on the same MCPS photometry used here, but
concentrating on the photometry of the region's abundant supergiant
population to provide higher temporal and spatial resolution in the
recent SFH For example, we were able to distinguish two bursts of
activity at 10~Myr and 30~Myr, whereas in the present analysis
(Figure~\ref{fig:sfh_consIII}), we are only able to resolve a single
recent burst.  The solution in Figure~\ref{fig:sfh_consIII} is
entirely consistent with the analysis of \cite{hz08}.  The SFH of this
region reveals that this is a relatively young structure in the LMC
disk: prior to 50~Myr ago, the SFH of this region is
indistinguishable from that of the void region
(Figure~\ref{fig:sfh_void}).

% SFH of Constellation III

\subsubsection{30~Doradus}\label{sec:30dor}

\begin{figure}[htbp]
\epsscale{0.85}
\plotone{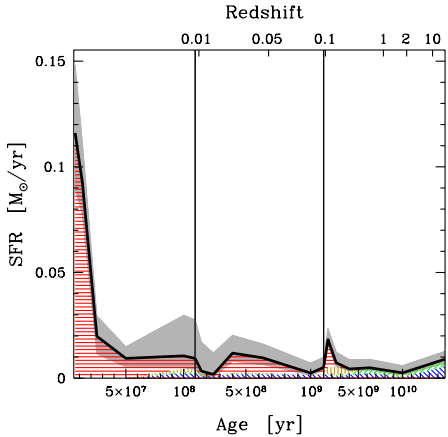}
\caption{The integrated SFH solution for the 30~Doradus region.  The
  shading and colors are the same as in
  Figure~\ref{fig:sfh_lmc}. \label{fig:sfh_30dor} }
\end{figure}

30~Doradus is one of the most active star-forming regions in the local
universe, and as such there is a lot of interest in understanding the
conditions that led to its formation.  Clues might be present in the
detailed, spatially-resolved reconstruction of its SFH.  The
integrated SFH of 30~Doradus (Figure~\ref{fig:sfh_30dor}) reveals
little more than the quintessential feature of 30~Doradus: that it is
a hotspot of very recent star-formation activity.
Figure~\ref{fig:sfhmap} shows that the 30~Doradus region is not
distinguishable in the SFH map until about 12~Myr ago, implying that,
like Constellation~III, 30~Doradus formed only very recently.  A more
targeted analysis that can achieve a higher age resolution for recent
ages (such as we have already done for Constellation~III;
\citealt{hz08}) might reveal more about the evolution of this
important star-forming region but is beyond the scope of this study.

% SFH of 30 Doradus

\subsubsection{The ``Northwest Void''}\label{sec:void}

\begin{figure}[htbp]
\epsscale{0.85}
\plotone{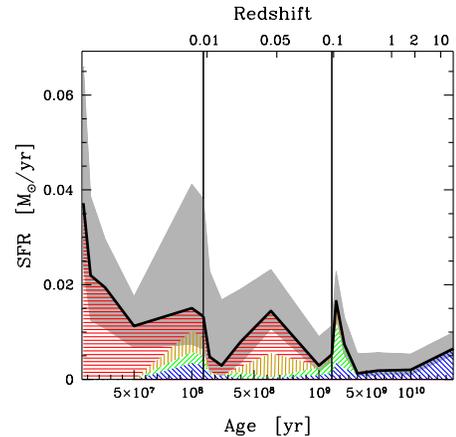}
\caption{The integrated SFH solution for the Northwest Void region.
  The shading and colors are the same as in
  Figure~\ref{fig:sfh_lmc}. \label{fig:sfh_void} }
\end{figure}

There is a large region in the northwestern disk that shows no
enhanced star formation activity at any epoch (see
Figure~\ref{fig:sfhmap}).  We refer to the region as the ``northwest
void''.  However, we see that the integrated SFH of the northwest void
(Figure~\ref{fig:sfh_void}) closely resembles the total integrated SFH
of the entire LMC (Figure~\ref{fig:sfh_lmc}).  In some sense, then,
the void offers a look at the LMC's typical stellar population,
unpolluted by recent localized deviations from the average behavior.

% SFH of the Northwest Void
                                                                                                                           
\subsection{Imprints of the LMC's Interaction History}\label{sec:interactions}

Until recently, the Magellanic Clouds were widely regarded as
long-term satellite galaxies of the Milky Way, in decaying orbits that
would eventually lead to their consumption by our Galaxy \citep{gar94,
  bek04}.  In \cite{hz04} and \cite{zh04}, we used this context to interpret the SFH
solution for the SMC; in particular, we noted coincidences between
peaks in the SMC's SFR and past perigalactic interactions with the LMC
and Milky Way.  Among these coincident peaks, the most striking was
one that occurred 2.5~Gyr ago, coinciding with what was previously
identified as the time of a triple-interaction
among the LMC, SMC,  and Milky Way.  We see the same 
peak in the
present analysis of the LMC's SFH,
further evidence for a major interaction at that time in their shared
history.

However, the idea that the Magellanic Clouds have always been satellites of the
Milky Way has been seriously challenged by a new measurement of the
LMC's proper motion, using differential astrometry against background
quasars from HST images \citep{kal06,bes07}.  The new proper motion is
twice as large as previous 
measurements. With this large a proper motion,
a closed orbit for the LMC around the Milky Way is 
possible only if 
the true mass of the Milky Way is larger than 
current measurements suggest.  In fact, the orbital reconstructions in \cite{bes07}
suggest that the LMC may not even be bound to the Local Group
potential. In their best-fit reconstruction, the LMC was over a Mpc away from
the Milky Way 5~Gyr ago.

\begin{figure}[htbp]
\epsscale{0.85}
\plotone{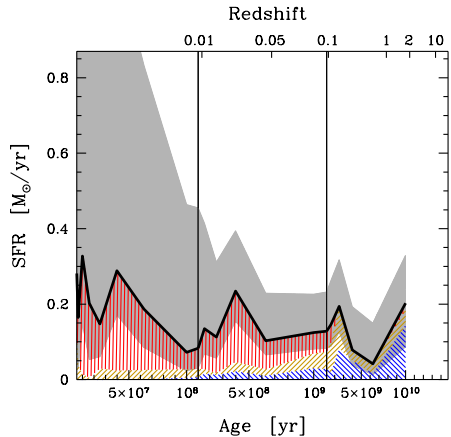}
\caption{The integrated SFH solution for the Small Magellanic Cloud,
  adapted from \cite{hz04}.  The shading and colors are the same as in
  Figure~\ref{fig:sfh_lmc}. \label{fig:sfh_smc} }
\end{figure}

The new proper motion measurement challenges the idea that the Clouds
are satellites of the Milky Way, but its impact on their gravitational
relationship with each other is much less clear.  Given the uncertainties
still present in the SMC proper motion measurement, It is still
possible that the Clouds are a long-term binary galaxy.
Assuming that past interactions between the Clouds 
trigger enhanced star formation
activity in both galaxies, we search for evidence of past
interactions between the Clouds by looking for coincident peaks in
their SFH solutions.  The SMC's integrated SFH solution is shown in
Figure~\ref{fig:sfh_smc}, adapted from \cite{hz04}.  Comparing the
SMC's solution to that of the LMC (Figure~\ref{fig:sfh_lmc}),
we do see two coincident peaks at 400~Myr and 2.5~Gyr which
suggest that the Clouds may indeed have a common interaction
history that extends back at least a few Gyr (while the upturn in the SFR in
both galaxies at about 5 Gyr ago, suggests their association may even extend
further back in time).

% SFH of the SMC

\subsection{The LMC's Chemical Enrichment History}\label{sec:ceh}

\begin{figure}[htbp]
\plotone{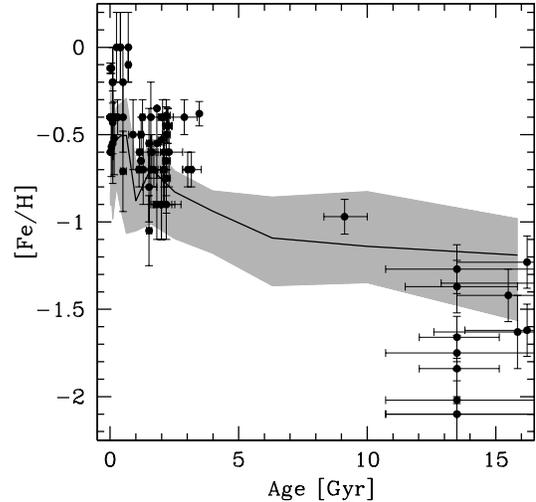}
\caption{The Age-Metallicity Relation for the LMC.  The points with
  errorbars are 85 LMC star clusters which have age and metallicity
  measurements in the literature.  The mean metallicity as a function
  of age derived from our SFH analysis is shown as a heavy line, and
  its statistical variance is shown as a grey envelope.  Our analysis
  contains only a single value for ages older than 4~Gyr, because we
  anchored the old SFH to published results based on deep HST
  imaging. \label{fig:agez} }
\end{figure}

The chemical enrichment history (CEH) of a galaxy is the complement of
its SFH, completing our understanding of the feedback cycles of star
formation and stellar evolution.  Over the past few decades, nearly a
hundred LMC star clusters have had both an accurate age estimate from
isochrone fitting to their CMDs, and a metallicity measurement from
either spectroscopy or isochrone fitting.  The resulting
age-metallicity relation (AMR) for LMC star clusters provides perhaps
the best available information on its CEH.  Several constructions of
the LMC's cluster AMR exist, based on the best values in the
literature at the time \citep{osm96, dir00}.  We present an updated
AMR in Figure~\ref{fig:agez}, consisting of age and metallicity values
of 85 LMC star clusters (the age and metallicity data were compiled
from a literature search, and are tabulated in Appendix~A).  This is 
not necessarily a complete sample, nor has it been vetted to necessarily 
present the highest quality measurements. As one example of the latter, a recent
study of four clusters \citep{mucc} presents spectroscopic abundance
measurements for three of the clusters in Appendix~A. The differences
in the abundance measurements as presented in the new study and
our tabulated literature search are $-$0.23, 0.00, and $-$0.09. While
at least the largest of these differences would be problematic for certain
issues, a quick inspection of Figure~\ref{fig:agez} reveals that both the 
scatter among clusters and the uncertainties in our measured chemical
evolution render such difference irrelevant for the purpose at hand.

% Age-metallicity relation derived from the SFH

The
essential features of the AMR have not changed: there is still a long
``Age Gap'', a dearth of cluster formation interrupted only by the
enigmatic ESO121-SC03.  Prior to the Age Gap, the LMC formed a number
of low-metallicity populous clusters, analogous to the Milky Way's
globular cluster population.  Following the Age Gap, cluster formation
resumed in the LMC, and proceeded roughly continuously
to the present day.  The clusters formed following the Age Gap have a
wide dispersion in metallicity, but they are significantly more
enriched than the ancient cluster population, and there is some
evidence for further enrichment with time for ages $<$4~Gyr.

Recently, \cite{car08} have examined the AMR for field stellar
populations, based on Calcium triplet spectroscopy of individual red
giants in four LMC fields.  They find that the disk AMR is similar to
that of the LMC star clusters: there was rapid initial enrichment to
[Fe/H]=$-$0.7 by about 10~Gyr ago.  Following this, the metallicity
increased only slowly until 4~Gyr ago, when it began a more rapid
increase to its present-day value of [Fe/H]=$-$0.2.  While
\citeauthor{car08} do see some field stars with ages spanning the
cluster Age Gap, their age uncertainties are necessarily large for
individual stars, and their reconstructed SFH (for their inner disk
fields) still indicates a quiescent epoch between 5 and 10~Gyr ago.
These conclusions differ in detail from those derived from the AMR
of the central region of the LMC \citep{cole05}, but the observed
AMR are entirely consistent.

Detailed abundance measurements of individual LMC stars also show
a chemical signature of this long quiescent epoch in the SFH
\citep{tsu95, hill00, pom08}.  The alpha-element abundances in these
stars indicates that enrichment by Type~Ia supernovae has been more
important in the LMC than in the Milky Way, which implies that the
cluster Age Gap was a general lull in the LMC's star formation rate,
and not simply the result of dynamical cluster destruction processes.

Into this context, we insert our reconstruction of the LMC's CEH,
based on the best-fit metallicities as a function of age in our
StarFISH analysis.  Photometry is an extremely blunt instrument with
which to determine metallicities, even in a statistical analysis
involving millions of stars.  However, the best-fit SFH solutions show
the bulk enrichment trend that is expected from the cyclical feedback
processes of star formation and stellar evolution. This congruence
with expectations gives us some confidence in at least the broad
outlines of the chemical enrichment history encoded in our solutions.

We use the stellar mass formed during each age bin and at each
metallicity to compute the stellar-mass weighted average metallicity
as a function of age.  This global mean age-metallicity relation is
shown in Figure~\ref{fig:agez}, overplotted with the LMC's cluster
AMR.  We assign uncertainties to our metallicity estimates based on
the difference between adjacent metallicity bins in our analysis and
take symmetric error bars for bins on the edges of our range. 
Because we have tied the old SFH to the HST results, we have
only a single best-fit metallicity for ages older than $\sim$4~Gyr.
Our reconstruction of the LMC's CEH is qualitatively similar to the
one we derived for the SMC \citep[see][]{hz04}.  In both
galaxies, the mean metallicity remained at a constant low value, until
about 4~Gyr ago when the metallicity began a steady ascent to the
present-day value.  In both galaxies, the age-metallicity relation
from our analysis is consistent with that of the star clusters, and
with the ``bursting'' enrichment model proposed by \cite{pt99}.
Interestingly, the enrichment trend in the LMC appears to have two
small interruptions at 1~Gyr and 100~Myr ago, but these episodes are
only marginally significant.

\section{Summary}\label{sec:summary}

In this paper, we present the first global reconstruction of the SFH
of the LMC, and indeed, the first global SFH solution for any galaxy
as large as 0.1~$L^\star$.  The SFH revealed by our solution is
largely consistent with the LMC's cluster formation history. There was
an initial epoch of star formation during which a significant portion
of the LMC's stellar mass was formed.  This era was followed by a long
quiescent epoch, during which star formation was suppressed throughout
the LMC.  Then, about 5~Gyr ago, star formation resumed throughout the
LMC, and has been ongoing since.  The global resumption of star
formation is strongly suggestive of a dramatic change in the LMC 5~Gyr
ago, such as the merger of a gas-rich dwarf galaxy or a particularly
dramatic tidal encounter, probably with the SMC given the similar
resumption of star formation seen in that galaxy's SFH.

The chemical enrichment history inferred from our analysis is also
consistent with the cluster age-metallicity relation. The early
history was characterized by rapid enrichment, reaching [Fe/H]=-1.2 by
about 12~Gyr ago.  The metallicity then remained stagnant throughout
the long quiescent epoch. When star formation resumed 5~Gyr ago,
the chemical abundances began a steady ascent to the present-day value
of [Fe/H]=-0.5.  The tight coupling between the field and cluster
populations in the LMC, in both their star formation and chemical
enrichment histories, argues for a strong connection between field and
cluster ``modes'' of star formation in this galaxy.

Since star formation resumed 5~Gyr ago, the mean star formation rate
has been approximately 0.2~$M_\odot$~yr$^{-1}$, modulated by
variations at the factor-of-two level.  We note that our measured
variations should be regarded as a lower limit, since strong
deviations in the star formation rate can be undetected by our
analysis if they are of sufficiently short duration.  We measure the
total number of stars formed during an epoch, we cannot constrain
whether the rate of star formation within that epoch was constant or
highly variable. 

There are several temporally coincident features in the global SFH solutions of
the LMC and SMC.  Both galaxies experienced a long quiescent epoch
starting roughly 10~Gyr ago, and ending 5~Gyr ago in the LMC, and
3~Gyr ago in the SMC.  Both galaxies experienced a significant peak in
the star formation rate 2--3~Gyr ago.  In the LMC, the stars that were
formed during this event are now centrally concentrated in the LMC,
although they don't seem to conform to the contours of the LMC bar.
In the SMC, the stars formed at this time are not centrally
concentrated today; indeed, they seem to be distributed in a ring-like
structure.  Both galaxies also show enhanced star-formation activity
around 400~Myr ago.  These correlations suggest a common history for
the Magellanic Clouds, stretching back over at least the past few~Gyr.

We examine the spatial variation of the LMC's SFH by comparing the
integrated solutions for eight subregions in the SFH map.  The SFH of
the LMC bar is qualitatively similar to that of non-bar regions,
indicating that the stars in the bar have likely always been part of
the LMC The SFHs of spiral-like ``arms'' extending off opposite ends
of the bar are strikingly dissimilar.  For example, the activity at
2.5~Gyr is very pronounced in the southeast arm, but is not
distinguishable in the northwest arm.  Also, the recent star formation
rate peaked 600~Myr ago in the southeast arm, whereas the star
formation rate in the northwest arm has climbed steadily over the past
Gyr.  The lack of symmetry in the histories of these ``arm''
structures suggests that these structures should not be regarded as a
spiral-density wave phenomenon in the disk of the LMC.  In a future
paper, we will examine the spatial relationships of events in the
LMC's recent SFH in greater detail, with an eye toward revealing some
of the physical processes driving star formation in the LMC disk.

The global SFH solution presented in this paper provides us with a
unique opportunity to study galaxy evolution from a novel angle.  Much
of what we have learned about galaxy evolution comes from the study of
large samples of galaxies at high redshift \citep[cf.][]{lilly, madau}.  In these studies,
evolution is necessarily inferential, because no single galaxy can be
followed through its history; we have only a series of ``snapshots''
showing the evolutionary state of populations of galaxies at a
particular point in their history.  The total SFH solution of the LMC
represents a nearly ideal complement to the body of work on the
evolution of galaxy populations at different redshifts: for a single
galaxy, at least, we now possess the knowledge of its entire past
history.  Other dwarf galaxies have had complete SFH solutions, but
the LMC is the first one that is massive enough to be 
well represented in large galaxy surveys out to at least $z=1$.  In a
follow-up paper, we will place the LMC in its cosmological context by
examining the evolution of its bulk properties, and by placing the
evolutionary track of the LMC on well-known examinations of galaxy
evolution from the galaxy-survey literature (such as the Madau-Lilly plot,
the galaxy color-magnitude diagram, and galaxy scaling relations \citep{zzg}).

\vskip 1in
\noindent 
Acknowledgments: We are grateful to Sean Points and Chris Smith for
providing us with the MCELS H$\alpha$ image of the LMC in advance of
its publication.  This work was partially supported by NASA through
Hubble Fellowship grant HF-01160.01-A awarded by the Space Telescope
Science Institute, which is operated by the Association of
Universities for Research in Astronomy, Inc., under NASA contract NAS
5-26555.  DZ acknowledges financial support from National Science
Foundation CAREER grant AST-973311, AST-0307482, NASA LTSA award
NNG05GE82G, and a fellowship from the David and Lucile Packard
Foundation.

\bibliographystyle{apj}
\bibliography{zaritsky}

\begin{thebibliography}{79}
\expandafter\ifx\csname natexlab\endcsname\relax\def\natexlab#1{#1}\fi

\bibitem[{{Alcock} {et~al.}(1999){Alcock}, {Allsman}, {Alves}, {Axelrod},
  {Becker}, {Bennett}, {Bersier}, {Cook}, {Freeman}, {Griest}, {Guern},
  {Lehner}, {Marshall}, {Minniti}, {Peterson}, {Pratt}, {Quinn}, {Rodgers},
  {Stubbs}, {Sutherland}, {Tomaney}, {Vandehei}, \& {Welch}}]{alc99}
{Alcock}, C., {Allsman}, R.~A., {Alves}, D.~R., {Axelrod}, T.~S., {Becker},
  A.~C., {Bennett}, D.~P., {Bersier}, D.~F., {Cook}, K.~H., {Freeman}, K.~C.,
  {Griest}, K., {Guern}, J.~A., {Lehner}, M., {Marshall}, S.~L., {Minniti}, D.,
  {Peterson}, B.~A., {Pratt}, M.~R., {Quinn}, P.~J., {Rodgers}, A.~W.,
  {Stubbs}, C.~W., {Sutherland}, W., {Tomaney}, A., {Vandehei}, T., \& {Welch},
  D.~L. 1999, \aj, 117, 920

\bibitem[{{Bekki} {et~al.}(2004){Bekki}, {Couch}, {Beasley}, {Forbes}, {Chiba},
  \& {Da Costa}}]{bek04}
{Bekki}, K., {Couch}, W.~J., {Beasley}, M.~A., {Forbes}, D.~A., {Chiba}, M., \&
  {Da Costa}, G.~S. 2004, \apjl, 610, L93

\bibitem[{{Bertelli} {et~al.}(1992){Bertelli}, {Mateo}, {Chiosi}, \&
  {Bressan}}]{ber92}
{Bertelli}, G., {Mateo}, M., {Chiosi}, C., \& {Bressan}, A. 1992, \apj, 388,
  400

\bibitem[{{Besla} {et~al.}(2007){Besla}, {Kallivayalil}, {Hernquist},
  {Robertson}, {Cox}, {van der Marel}, \& {Alcock}}]{bes07}
{Besla}, G., {Kallivayalil}, N., {Hernquist}, L., {Robertson}, B., {Cox},
  T.~J., {van der Marel}, R.~P., \& {Alcock}, C. 2007, \apj, 668, 949

\bibitem[{{Bica} {et~al.}(1998){Bica}, {Geisler}, {Dottori}, {Clari{\'a}},
  {Piatti}, \& {Santos}}]{bic98}
{Bica}, E., {Geisler}, D., {Dottori}, H., {Clari{\'a}}, J.~J., {Piatti}, A.~E.,
  \& {Santos}, Jr., J.~F.~C. 1998, \aj, 116, 723

\bibitem[{{Brocato} {et~al.}(1996){Brocato}, {Castellani}, {Ferraro},
  {Piersimoni}, \& {Testa}}]{bro96}
{Brocato}, E., {Castellani}, V., {Ferraro}, F.~R., {Piersimoni}, A.~M., \&
  {Testa}, V. 1996, \mnras, 282, 614

\bibitem[{{Carrera} {et~al.}(2008){Carrera}, {Gallart}, {Hardy}, {Aparicio}, \&
  {Zinn}}]{car08}
{Carrera}, R., {Gallart}, C., {Hardy}, E., {Aparicio}, A., \& {Zinn}, R. 2008,
  \aj, 135, 836

\bibitem[{{Cole} {et~al.}(2005){Cole}, {Tolstoy}, {Gallagher}, \&
  {Smecker-Hane}}]{cole05}
{Cole}, A.~A., {Tolstoy}, E., {Gallagher}, III, J.~S., \& {Smecker-Hane}, T.~A.
  2005, \aj, 129, 1465

\bibitem[{{Da Costa}(1991)}]{dc91}
{Da Costa}, G.~S. 1991, in IAU Symp. 148: The Magellanic Clouds, 183--+

\bibitem[{{Da Costa} \& {Hatzidimitriou}(1998)}]{dh98}
{Da Costa}, G.~S. \& {Hatzidimitriou}, D. 1998, \aj, 115, 1934

\bibitem[{{Dirsch} {et~al.}(2000){Dirsch}, {Richtler}, {Gieren}, \&
  {Hilker}}]{dir00}
{Dirsch}, B., {Richtler}, T., {Gieren}, W.~P., \& {Hilker}, M. 2000, \aap, 360,
  133

\bibitem[{{Dolphin}(2000)}]{dol00}
{Dolphin}, A.~E. 2000, \mnras, 313, 281

\bibitem[{{Dolphin}(2002)}]{dol02}
---. 2002, \mnras, 332, 91

\bibitem[{{Ferraro} {et~al.}(2006){Ferraro}, {Mucciarelli}, {Carretta}, \&
  {Origlia}}]{fer06}
{Ferraro}, F.~R., {Mucciarelli}, A., {Carretta}, E., \& {Origlia}, L. 2006,
  \apjl, 645, L33

\bibitem[{{Gallagher} {et~al.}(1996){Gallagher}, {Mould}, {de Feijter},
  {Holtzman}, {Stappers}, {Watson}, {Trauger}, {Ballester}, {Burrows},
  {Casertano}, {Clarke}, {Crisp}, {Griffiths}, {Hester}, {Hoessel}, {Krist},
  {Matthews}, {Scowen}, {Stapelfeld}, \& {Westphal}}]{glg96}
{Gallagher}, J.~S., {Mould}, J.~R., {de Feijter}, E., {Holtzman}, J.,
  {Stappers}, B., {Watson}, A., {Trauger}, J., {Ballester}, G.~E., {Burrows},
  C.~J., {Casertano}, S., {Clarke}, J.~T., {Crisp}, D., {Griffiths}, R.~E.,
  {Hester}, J.~J., {Hoessel}, J., {Krist}, J., {Matthews}, L.~D., {Scowen},
  P.~A., {Stapelfeld}, K.~R., \& {Westphal}, J.~A. 1996, \apj, 466, 732

\bibitem[{{Gardiner} {et~al.}(1994){Gardiner}, {Sawa}, \& {Fujimoto}}]{gar94}
{Gardiner}, L.~T., {Sawa}, T., \& {Fujimoto}, M. 1994, \mnras, 266, 567

\bibitem[{{Geha} {et~al.}(1998){Geha}, {Holtzman}, {Mould}, {Gallagher},
  {Watson}, {Cole}, {Grillmair}, {Stapelfeldt}, {Ballester}, {Burrows},
  {Clarke}, {Crisp}, {Evans}, {Griffiths}, {Hester}, {Scowen}, {Trauger}, \&
  {Westphal}}]{geh98}
{Geha}, M.~C., {Holtzman}, J.~A., {Mould}, J.~R., {Gallagher}, III, J.~S.,
  {Watson}, A.~M., {Cole}, A.~A., {Grillmair}, C.~J., {Stapelfeldt}, K.~R.,
  {Ballester}, G.~E., {Burrows}, C.~J., {Clarke}, J.~T., {Crisp}, D., {Evans},
  R.~W., {Griffiths}, R.~E., {Hester}, J.~J., {Scowen}, P.~A., {Trauger},
  J.~T., \& {Westphal}, J.~A. 1998, \aj, 115, 1045

\bibitem[{{Geisler} {et~al.}(2003){Geisler}, {Piatti}, {Bica}, \&
  {Clari{\'a}}}]{gei03}
{Geisler}, D., {Piatti}, A.~E., {Bica}, E., \& {Clari{\'a}}, J.~J. 2003,
  \mnras, 341, 771

\bibitem[{{Gilmozzi} {et~al.}(1994){Gilmozzi}, {Kinney}, {Ewald}, {Panagia}, \&
  {Romaniello}}]{gil94}
{Gilmozzi}, R., {Kinney}, E.~K., {Ewald}, S.~P., {Panagia}, N., \&
  {Romaniello}, M. 1994, \apjl, 435, L43

\bibitem[{{Girardi} {et~al.}(2002){Girardi}, {Bertelli}, {Bressan}, {Chiosi},
  {Groenewegen}, {Marigo}, {Salasnich}, \& {Weiss}}]{gir02}
{Girardi}, L., {Bertelli}, G., {Bressan}, A., {Chiosi}, C., {Groenewegen},
  M.~A.~T., {Marigo}, P., {Salasnich}, B., \& {Weiss}, A. 2002, \aap, 391, 195

\bibitem[{{Grocholski} {et~al.}(2006){Grocholski}, {Cole}, {Sarajedini},
  {Geisler}, \& {Smith}}]{gro06}
{Grocholski}, A.~J., {Cole}, A.~A., {Sarajedini}, A., {Geisler}, D., \&
  {Smith}, V.~V. 2006, \aj, 132, 1630

\bibitem[{{Hardy} {et~al.}(1984){Hardy}, {Buonanno}, {Corsi}, {Janes}, \&
  {Schommer}}]{hardy}
{Hardy}, E., {Buonanno}, R., {Corsi}, C.~E., {Janes}, K.~A., \& {Schommer},
  R.~A. 1984, \apj, 278, 592

\bibitem[{{Harris} \& {Zaritsky}(2001)}]{hz01}
{Harris}, J. \& {Zaritsky}, D. 2001, \apjs, 136, 25

\bibitem[{{Harris} \& {Zaritsky}(2004)}]{hz04}
---. 2004, \aj, 127, 1531

\bibitem[{{Harris} \& {Zaritsky}(2008)}]{hz08}
---. 2008, accepted for publication in {\it PASA}

\bibitem[{{Hilker} {et~al.}(1995){Hilker}, {Richtler}, \& {Gieren}}]{hil95}
{Hilker}, M., {Richtler}, T., \& {Gieren}, W. 1995, \aap, 294, 648

\bibitem[{{Hill} {et~al.}(1995){Hill}, {Cheng}, {Bohlin}, {O'Connell},
  {Roberts}, {Smith}, \& {Stecher}}]{hill95}
{Hill}, R.~S., {Cheng}, K.-P., {Bohlin}, R.~C., {O'Connell}, R.~W., {Roberts},
  M.~S., {Smith}, A.~M., \& {Stecher}, T.~P. 1995, \apj, 446, 622

\bibitem[{{Hill} {et~al.}(2000){Hill}, {Fran{\c c}ois}, {Spite}, {Primas}, \&
  {Spite}}]{hill00}
{Hill}, V., {Fran{\c c}ois}, P., {Spite}, M., {Primas}, F., \& {Spite}, F.
  2000, \aap, 364, L19

\bibitem[{{Hodge}(1960)}]{hod60}
{Hodge}, P.~W. 1960, \apj, 131, 351

\bibitem[{{Hodge}(1961)}]{hod61}
---. 1961, \apj, 133, 413

\bibitem[{{Holtzman} {et~al.}(1999){Holtzman}, {Gallagher}, {Cole}, {Mould},
  {Grillmair}, {Ballester}, {Burrows}, {Clarke}, {Crisp}, {Evans}, {Griffiths},
  {Hester}, {Hoessel}, {Scowen}, {Stapelfeldt}, {Trauger}, \& {Watson}}]{hol99}
{Holtzman}, J.~A., {Gallagher}, J.~S., I., {Cole}, A.~A., {Mould}, J.~R.,
  {Grillmair}, C.~J., {Ballester}, G.~E., {Burrows}, C.~J., {Clarke}, J.~T.,
  {Crisp}, D., {Evans}, R.~W., {Griffiths}, R.~E., {Hester}, J.~J., {Hoessel},
  J.~G., {Scowen}, P.~A., {Stapelfeldt}, K.~R., {Trauger}, J.~T., \& {Watson},
  A.~M. 1999, \aj, 118, 2262

\bibitem[{{Hunter} {et~al.}(1995){Hunter}, {Shaya}, {Holtzman}, {Light},
  {O'Neil}, \& {Lynds}}]{hun95}
{Hunter}, D.~A., {Shaya}, E.~J., {Holtzman}, J.~A., {Light}, R.~M., {O'Neil},
  Jr., E.~J., \& {Lynds}, R. 1995, \apj, 448, 179

\bibitem[{{Johnson} {et~al.}(1999){Johnson}, {Bolte}, {Stetson}, {Hesser}, \&
  {Somerville}}]{joh99}
{Johnson}, J.~A., {Bolte}, M., {Stetson}, P.~B., {Hesser}, J.~E., \&
  {Somerville}, R.~S. 1999, \apj, 527, 199

\bibitem[{{Kallivayalil} {et~al.}(2006){Kallivayalil}, {van der Marel},
  {Alcock}, {Axelrod}, {Cook}, {Drake}, \& {Geha}}]{kal06}
{Kallivayalil}, N., {van der Marel}, R.~P., {Alcock}, C., {Axelrod}, T.,
  {Cook}, K.~H., {Drake}, A.~J., \& {Geha}, M. 2006, \apj, 638, 772

\bibitem[{{Kerber} {et~al.}(2007){Kerber}, {Santiago}, \& {Brocato}}]{ker07}
{Kerber}, L.~O., {Santiago}, B.~X., \& {Brocato}, E. 2007, \aap, 462, 139

\bibitem[{{Kim} {et~al.}(1998){Kim}, {Staveley-Smith}, {Dopita}, {Freeman},
  {Sault}, {Kesteven}, \& {McConnell}}]{kim98}
{Kim}, S., {Staveley-Smith}, L., {Dopita}, M.~A., {Freeman}, K.~C., {Sault},
  R.~J., {Kesteven}, M.~J., \& {McConnell}, D. 1998, \apj, 503, 674

\bibitem[{{Lilly} {et~al.}(1996){Lilly}, {Le Fevre}, {Hammer}, \&
  {Crampton}}]{lilly}
{Lilly}, S.~J., {Le Fevre}, O., {Hammer}, F., \& {Crampton}, D. 1996, \apjl,
  460, L1+

\bibitem[{{Mackey} \& {Broby Nielsen}(2007)}]{mac07}
{Mackey}, A.~D. \& {Broby Nielsen}, P. 2007, \mnras, 379, 151

\bibitem[{{Mackey} \& {Gilmore}(2004)}]{mac04}
{Mackey}, A.~D. \& {Gilmore}, G.~F. 2004, \mnras, 352, 153

\bibitem[{{Mackey} {et~al.}(2006){Mackey}, {Payne}, \& {Gilmore}}]{mac06}
{Mackey}, A.~D., {Payne}, M.~J., \& {Gilmore}, G.~F. 2006, \mnras, 369, 921

\bibitem[{{Madau} {et~al.}(1996){Madau}, {Ferguson}, {Dickinson}, {Giavalisco},
  {Steidel}, \& {Fruchter}}]{madau}
{Madau}, P., {Ferguson}, H.~C., {Dickinson}, M.~E., {Giavalisco}, M.,
  {Steidel}, C.~C., \& {Fruchter}, A. 1996, \mnras, 283, 1388

\bibitem[{{Meixner} {et~al.}(2006){Meixner}, {Gordon}, {Indebetouw}, {Hora},
  {Whitney}, {Blum}, {Reach}, {Bernard}, {Meade}, {Babler}, {Engelbracht},
  {For}, {Misselt}, {Vijh}, {Leitherer}, {Cohen}, {Churchwell}, {Boulanger},
  {Frogel}, {Fukui}, {Gallagher}, {Gorjian}, {Harris}, {Kelly}, {Kawamura},
  {Kim}, {Latter}, {Madden}, {Markwick-Kemper}, {Mizuno}, {Mizuno}, {Mould},
  {Nota}, {Oey}, {Olsen}, {Onishi}, {Paladini}, {Panagia}, {Perez-Gonzalez},
  {Shibai}, {Sato}, {Smith}, {Staveley-Smith}, {Tielens}, {Ueta}, {Dyk},
  {Volk}, {Werner}, \& {Zaritsky}}]{mei06}
{Meixner}, M., {Gordon}, K.~D., {Indebetouw}, R., {Hora}, J.~L., {Whitney}, B.,
  {Blum}, R., {Reach}, W., {Bernard}, J.-P., {Meade}, M., {Babler}, B.,
  {Engelbracht}, C.~W., {For}, B.-Q., {Misselt}, K., {Vijh}, U., {Leitherer},
  C., {Cohen}, M., {Churchwell}, E.~B., {Boulanger}, F., {Frogel}, J.~A.,
  {Fukui}, Y., {Gallagher}, J., {Gorjian}, V., {Harris}, J., {Kelly}, D.,
  {Kawamura}, A., {Kim}, S., {Latter}, W.~B., {Madden}, S., {Markwick-Kemper},
  C., {Mizuno}, A., {Mizuno}, N., {Mould}, J., {Nota}, A., {Oey}, M.~S.,
  {Olsen}, K., {Onishi}, T., {Paladini}, R., {Panagia}, N., {Perez-Gonzalez},
  P., {Shibai}, H., {Sato}, S., {Smith}, L., {Staveley-Smith}, L., {Tielens},
  A.~G.~G.~M., {Ueta}, T., {Dyk}, S.~V., {Volk}, K., {Werner}, M., \&
  {Zaritsky}, D. 2006, \aj, 132, 2268

\bibitem[{{Mizuno} {et~al.}(2001){Mizuno}, {Yamaguchi}, {Mizuno}, {Rubio},
  {Abe}, {Saito}, {Onishi}, {Yonekura}, {Yamaguchi}, {Ogawa}, \&
  {Fukui}}]{miz01}
{Mizuno}, N., {Yamaguchi}, R., {Mizuno}, A., {Rubio}, M., {Abe}, R., {Saito},
  H., {Onishi}, T., {Yonekura}, Y., {Yamaguchi}, N., {Ogawa}, H., \& {Fukui},
  Y. 2001, \pasj, 53, 971

\bibitem[{{Mucciarelli} {et~al.}(2008){Mucciarelli}, {Carretta}, {Origlia}, \&
  {Ferraro}}]{mucc}
{Mucciarelli}, A., {Carretta}, E., {Origlia}, L., \& {Ferraro}, F.~R. 2008,
  \aj, 136, 375

\bibitem[{{Nikolaev} {et~al.}(2004){Nikolaev}, {Drake}, {Keller}, {Cook},
  {Dalal}, {Griest}, {Welch}, \& {Kanbur}}]{nik04}
{Nikolaev}, S., {Drake}, A.~J., {Keller}, S.~C., {Cook}, K.~H., {Dalal}, N.,
  {Griest}, K., {Welch}, D.~L., \& {Kanbur}, S.~M. 2004, \apj, 601, 260

\bibitem[{{Nikolaev} \& {Weinberg}(2000)}]{nw2000}
{Nikolaev}, S. \& {Weinberg}, M.~D. 2000, \apj, 542, 804

\bibitem[{{Oey} \& {Massey}(1995)}]{om95}
{Oey}, M.~S. \& {Massey}, P. 1995, \apj, 452, 210

\bibitem[{{Olsen}(1999)}]{ols99}
{Olsen}, K. A.~G. 1999, \aj, 117, 2244

\bibitem[{{Olsen} {et~al.}(1998){Olsen}, {Hodge}, {Mateo}, {Olszewski},
  {Schommer}, {Suntzeff}, \& {Walker}}]{ols98}
{Olsen}, K.~A.~G., {Hodge}, P.~W., {Mateo}, M., {Olszewski}, E.~W., {Schommer},
  R.~A., {Suntzeff}, N.~B., \& {Walker}, A.~R. 1998, \mnras, 300, 665

\bibitem[{{Olsen} {et~al.}(1997){Olsen}, {Hodge}, {Wilcots}, \&
  {Pastwick}}]{ols97}
{Olsen}, K.~A.~G., {Hodge}, P.~W., {Wilcots}, E.~M., \& {Pastwick}, L. 1997,
  \apj, 475, 545

\bibitem[{{Olszewski} {et~al.}(1996){Olszewski}, {Suntzeff}, \&
  {Mateo}}]{osm96}
{Olszewski}, E.~W., {Suntzeff}, N.~B., \& {Mateo}, M. 1996, \araa, 34, 511

\bibitem[{{Pagel} \& {Tautvaisien{\.e}}(1999)}]{pt99}
{Pagel}, B. E.~J. \& {Tautvaisien{\.e}}, G. 1999, \apss, 265, 461

\bibitem[{{Payne-Gaposhkin}(1972)}]{pg72}
{Payne-Gaposhkin}, C. 1972, in IAU Colloq. 17: Age des Etoiles, ed. G.~{Cayrel
  de Strobel} \& A.~M. {Delplace}, 3--+

\bibitem[{{Piatti} {et~al.}(2003){Piatti}, {Bica}, {Geisler}, \&
  {Clari{\'a}}}]{pia03}
{Piatti}, A.~E., {Bica}, E., {Geisler}, D., \& {Clari{\'a}}, J.~J. 2003,
  \mnras, 344, 965

\bibitem[{{Piatti} {et~al.}(2002){Piatti}, {Sarajedini}, {Geisler}, {Bica}, \&
  {Clari{\'a}}}]{pia02}
{Piatti}, A.~E., {Sarajedini}, A., {Geisler}, D., {Bica}, E., \& {Clari{\'a}},
  J.~J. 2002, \mnras, 329, 556

\bibitem[{{Points}(2008)}]{poi08}
{Points}, S. 2008, mCELS H$\alpha$ image, private communication

\bibitem[{{Pomp{\'e}ia} {et~al.}(2008){Pomp{\'e}ia}, {Hill}, {Spite}, {Cole},
  {Primas}, {Romaniello}, {Pasquini}, {Cioni}, \& {Smecker Hane}}]{pom08}
{Pomp{\'e}ia}, L., {Hill}, V., {Spite}, M., {Cole}, A., {Primas}, F.,
  {Romaniello}, M., {Pasquini}, L., {Cioni}, M.-R., \& {Smecker Hane}, T. 2008,
  \aap, 480, 379

\bibitem[{{Sagar} \& {Pandey}(1989)}]{sp89}
{Sagar}, R. \& {Pandey}, A.~K. 1989, \aaps, 79, 407

\bibitem[{{Sagar} \& {Richtler}(1991)}]{sr91}
{Sagar}, R. \& {Richtler}, T. 1991, \aap, 250, 324

\bibitem[{{Schinnerer} {et~al.}(2006){Schinnerer}, {B{\"o}ker}, {Emsellem}, \&
  {Lisenfeld}}]{sch06}
{Schinnerer}, E., {B{\"o}ker}, T., {Emsellem}, E., \& {Lisenfeld}, U. 2006,
  \apj, 649, 181

\bibitem[{{Schwarzschild}(1958)}]{sch58}
{Schwarzschild}, M. 1958, {Structure and evolution of the stars.} (Princeton,
  Princeton University Press, 1958.)

\bibitem[{{Smecker-Hane} {et~al.}(2002){Smecker-Hane}, {Cole}, {Gallagher}, \&
  {Stetson}}]{sh02}
{Smecker-Hane}, T.~A., {Cole}, A.~A., {Gallagher}, J.~S., \& {Stetson}, P.~B.
  2002, \apj, 566, 239

\bibitem[{{Staveley-Smith} {et~al.}(2003){Staveley-Smith}, {Kim}, {Calabretta},
  {Haynes}, \& {Kesteven}}]{ss03}
{Staveley-Smith}, L., {Kim}, S., {Calabretta}, M.~R., {Haynes}, R.~F., \&
  {Kesteven}, M.~J. 2003, \mnras, 339, 87

\bibitem[{{Tsujimoto} {et~al.}(1995){Tsujimoto}, {Nomoto}, {Yoshii},
  {Hashimoto}, {Yanagida}, \& {Thielemann}}]{tsu95}
{Tsujimoto}, T., {Nomoto}, K., {Yoshii}, Y., {Hashimoto}, M., {Yanagida}, S.,
  \& {Thielemann}, F.-K. 1995, \mnras, 277, 945

\bibitem[{{Vallenari} {et~al.}(1994{\natexlab{a}}){Vallenari}, {Aparicio},
  {Fagotto}, \& {Chiosi}}]{val94a}
{Vallenari}, A., {Aparicio}, A., {Fagotto}, F., \& {Chiosi}, C.
  1994{\natexlab{a}}, \aap, 284, 424

\bibitem[{{Vallenari} {et~al.}(1994{\natexlab{b}}){Vallenari}, {Aparicio},
  {Fagotto}, {Chiosi}, {Ortolani}, \& {Meylan}}]{val94b}
{Vallenari}, A., {Aparicio}, A., {Fagotto}, F., {Chiosi}, C., {Ortolani}, S.,
  \& {Meylan}, G. 1994{\natexlab{b}}, \aap, 284, 447

\bibitem[{{van den Bergh}(1981)}]{vdb81}
{van den Bergh}, S. 1981, \aaps, 46, 79

\bibitem[{{van der Marel} {et~al.}(2002){van der Marel}, {Alves}, {Hardy}, \&
  {Suntzeff}}]{vdm02}
{van der Marel}, R.~P., {Alves}, D.~R., {Hardy}, E., \& {Suntzeff}, N.~B. 2002,
  \aj, 124, 2639

\bibitem[{{van der Marel} \& {Cioni}(2001)}]{vdm01a}
{van der Marel}, R.~P. \& {Cioni}, M.-R.~L. 2001, \aj, 122, 1807

\bibitem[{{Will} {et~al.}(1996){Will}, {Bomans}, {Vallenari}, {Schmidt}, \& {de
  Boer}}]{wil96}
{Will}, J.-M., {Bomans}, D.~J., {Vallenari}, A., {Schmidt}, J.~H.~K., \& {de
  Boer}, K.~S. 1996, \aap, 315, 125

\bibitem[{{Woo} {et~al.}(2003){Woo}, {Gallart}, {Demarque}, {Yi}, \&
  {Zoccali}}]{woo03}
{Woo}, J.-H., {Gallart}, C., {Demarque}, P., {Yi}, S., \& {Zoccali}, M. 2003,
  \aj, 125, 754

\bibitem[{{Zaritsky}(1999)}]{zar99}
{Zaritsky}, D. 1999, \aj, 118, 2824

\bibitem[{{Zaritsky}(2004)}]{zar04b}
---. 2004, \apjl, 614, L37

\bibitem[{{Zaritsky} \& {Harris}(2004)}]{zh04}
{Zaritsky}, D. \& {Harris}, J. 2004, \apj, 604, 167

\bibitem[{{Zaritsky} {et~al.}(1997){Zaritsky}, {Harris}, \& {Thompson}}]{zar97}
{Zaritsky}, D., {Harris}, J., \& {Thompson}, I. 1997, \aj, 114, 1002

\bibitem[{{Zaritsky} {et~al.}(2004){Zaritsky}, {Harris}, {Thompson}, \&
  {Grebel}}]{zar04}
{Zaritsky}, D., {Harris}, J., {Thompson}, I.~B., \& {Grebel}, E.~K. 2004, \aj,
  128, 1606

\bibitem[{{Zaritsky} {et~al.}(2002){Zaritsky}, {Harris}, {Thompson}, {Grebel},
  \& {Massey}}]{zar02}
{Zaritsky}, D., {Harris}, J., {Thompson}, I.~B., {Grebel}, E.~K., \& {Massey},
  P. 2002, \aj, 123, 855

\bibitem[{{Zaritsky} {et~al.}(1996){Zaritsky}, {Schectman}, \&
  {Bredthauer}}]{zar96}
{Zaritsky}, D., {Schectman}, S.~A., \& {Bredthauer}, G. 1996, \pasp, 108, 104

\bibitem[{{Zaritsky} {et~al.}(2008){Zaritsky}, {Zabludoff}, \&
  {Gonzalez}}]{zzg}
{Zaritsky}, D., {Zabludoff}, A.~I., \& {Gonzalez}, A.~H. 2008, \apj, 682, 68

\end{thebibliography}

\clearpage

\appendix

\section{LMC Cluster Ages and Metallicities from the Literature}

Following \cite{dir00}, we have collected a set of age and metallicity
measurements of LMC stars clusters from the peer-review literature.
The compiled data for 85 star clusters in the LMC are presented in
Table~\ref{tab:cluster_agez}; these are the cluster data we used in
Figure~\ref{fig:agez}.  The cluster ages are derived from isochrone
fitting in a CMD plane, and the metallicities are derived from
spectroscopy or isochrone fitting.

%Table of LMC Cluster Ages and Metallicities from the literature

\begin{deluxetable*}{rrrrrrl}
\tablecaption{LMC Cluster Ages and Metallicities \label{tab:cluster_agez}}
\tablewidth{0pt}
\tablehead{
  \colhead{Cluster} &
  \colhead{[Fe/H]} & \colhead{$\sigma_{[Fe/H]}$} &
  \colhead{log(Age)} & \colhead{$\sigma_{log(Age)}$} &
  \colhead{Reference} & \colhead{Notes}
}
\startdata
Reticulum & $-$1.66 & 0.12 & 10.13 &  0.05 & \cite{mac04}  & {a}  \\
BRHTb     & $-$0.40 & 0.0  &  8.0  &  0.0  & \cite{pia03}  & {bc} \\
ESO121    & $-$0.97 & 0.1  &  9.96 &  0.04 & \cite{mac06}  & \nodata  \\
R136      & $-$0.4  & 0.0  &  7.0  &  0.0  & \cite{hun95}  & {bc} \\
OHSC33    & $-$1.05 & 0.2  &  9.18 &  0.03 & \cite{bic98}  & \nodata  \\
OHSC37    & $-$0.7  & 0.2  &  9.32 &  0.03 & \cite{bic98}  & \nodata  \\
Hodge11   & $-$1.84 & 0.0  & 10.13 &  0.05 & \cite{gro06}  & {ad} \\
LH47/48   & $-$0.4  & 0.0  &  6.3  &  0.0  & \cite{om95}   & {bc} \\
LH52/53   & $-$0.4  & 0.0  &  7.0  &  0.0  & \cite{hill95} & {bc} \\
LH72      & $-$0.6  & 0.0  &  6.95 &  0.25 & \cite{ols97}  & {b}  \\
LH77      & $-$0.4  & 0.0  &  7.20 &  0.14 & \cite{dh98}   & {b}  \\
LW224     &  0.0  & 0.2  &  8.85 &  0.0  & \cite{pia03}  & {c}  \\
SL8       & $-$0.55 & 0.2  &  9.26 &  0.03 & \cite{bic98}  & \nodata  \\
SL126     & $-$0.5  & 0.2  &  9.34 &  0.03 & \cite{bic98}  & \nodata  \\
SL218     & $-$0.40 & 0.0  &  7.70 &  0.0  & \cite{pia03}  & \nodata  \\
SL244     & $-$0.7  & 0.2  &  9.11 &  0.09 & \cite{gei03}  & {e}  \\
SL262     & $-$0.6  & 0.2  &  9.32 &  0.03 & \cite{bic98}  & \nodata  \\
SL359     & $-$0.4  & 0.2  &  9.20 &  0.1  & \cite{gei03}  & \nodata  \\
SL388     & $-$0.39 & 0.05 &  9.34 &  0.03 & \cite{gro06}  & {f}  \\
SL444     & $-$0.4  & 0.2  &  8.7  &  0.0  & \cite{pia03}  & {c}  \\
SL446A    & $-$0.9  & 0.2  &  9.34 &  0.1  & \cite{gei03}  & \nodata  \\
SL451     & $-$0.75 & 0.2  &  9.34 &  0.03 & \cite{bic98}  & \nodata  \\
SL503     & $-$0.4  & 0.0  &  7.20 &  0.22 & \cite{dh98}   & {b}  \\
SL505     & $-$0.5  & 0.2  &  8.95 &  0.09 & \cite{gei03}  & \nodata  \\
SL506     & $-$0.45 & 0.1  &  9.35 &  0.03 & \cite{ker07}  & \nodata  \\
SL509     & $-$0.65 & 0.0  &  9.08 &  0.0  & \cite{pia03}  & {bc} \\
SL548     &  0.0  & 0.2  &  8.60 &  0.0  & \cite{pia03}  & {c}  \\
SL549     & $-$0.9  & 0.2  &  9.30 &  0.1  & \cite{gei03}  & \nodata  \\
SL555     & $-$0.7  & 0.2  &  9.20 &  0.12 & \cite{gei03}  & \nodata  \\
SL556     & $-$0.7  & 0.2  &  9.32 &  0.0  & \cite{woo03}  & {c}  \\
SL663     & $-$0.7  & 0.1  &  9.50 &  0.05 & \cite{ker07}  & {e}  \\
SL674     & $-$0.9  & 0.2  &  9.30 &  0.08 & \cite{gei03}  & \nodata  \\
SL678     & $-$0.8  & 0.2  &  9.18 &  0.08 & \cite{gei03}  & \nodata  \\
SL769     & $-$0.35 & 0.0  &  9.26 &  0.0  & \cite{pia03}  & {bc} \\
SL817     & $-$0.55 & 0.2  &  9.18 &  0.03 & \cite{bic98}  & \nodata  \\             
SL842     & $-$0.65 & 0.2  &  9.34 &  0.03 & \cite{bic98}  & \nodata  \\
SL862     & $-$0.9  & 0.2  &  9.26 &  0.03 & \cite{bic98}  & \nodata  \\
SL896     & $-$0.6  & 0.2  &  9.36 &  0.09 & \cite{pia02}  & \nodata  \\
NGC1651   & $-$0.53 & 0.03 &  9.30 &  0.03 & \cite{gro06}  & {eg} \\
NGC1711   & $-$0.57 & 0.17 &  7.7  &  0.05 & \cite{dir00}  & \nodata  \\
NGC1718   & $-$0.4  & 0.1  &  9.31 &  0.03 & \cite{ker07}  & {e}  \\
NGC1754   & $-$1.42 & 0.15 & 10.19 &  0.06 & \cite{ols98}  & \nodata  \\
NGC1777   & $-$0.6  & 0.1  &  9.06 &  0.04 & \cite{ker07}  & \nodata  \\
NGC1786   & $-$2.1  & 0.3  & 10.13 &  0.1  & \cite{bro96}  & {h}  \\
NGC1806   & $-$0.71 & 0.23 &  8.7  &  0.1  & \cite{dir00}  & \nodata  \\
NGC1831   & $-$0.1  & 0.1  &  8.85 &  0.05 & \cite{ker07}  & \nodata  \\
NGC1835   & $-$1.62 & 0.15 & 10.21 &  0.07 & \cite{ols98}  & \nodata  \\
NGC1836   &  0.0  & 0.2  &  8.60 &  0.0  & \cite{pia03}  & {c}  \\
NGC1838   & $-$0.4  & 0.0  &  8.0  &  0.0  & \cite{pia03}  & {bc} \\
NGC1839   & $-$0.4  & 0.0  &  8.10 &  0.0  & \cite{pia03}  & {bc} \\
NGC1841   & $-$2.02 & 0.02 & 10.13 &  0.1  & \cite{gro06}  & {ai} \\
NGC1846   & $-$0.40 & 0.0  &  9.3  &  0.03 & \cite{mac07}  & {b}  \\
NGC1850A  & $-$0.12 & 0.03 &  7.7  &  0.1  & \cite{gil94}  & \nodata  \\
NGC1850B  & $-$0.12 & 0.03 &  6.6  &  0.1  & \cite{gil94}  & \nodata  \\
NGC1856   & $-$0.4  & 0.1  &  8.47 &  0.04 & \cite{ker07}  & \nodata  \\
NGC1858   & $-$0.4  & 0.0  &  6.9  &  0.0  & \cite{val94b} & {bc} \\
NGC1860   &  0.0  & 0.2  &  8.4  &  0.0  & \cite{pia03}  & {c}  \\
NGC1863   & $-$0.40 & 0.0  &  7.7  &  0.0  & \cite{pia03}  & {bc} \\
NGC1865   & $-$0.2  & 0.2  &  8.7  &  0.0  & \cite{pia03}  & {c}  \\
NGC1866   & $-$0.43 & 0.18 &  8.0  &  0.0  & \cite{hil95}  & {c}  \\
NGC1868   & $-$0.7  & 0.1  &  9.05 &  0.03 & \cite{ker07}  & \nodata  \\
NGC1898   & $-$1.37 & 0.15 & 10.13 &  0.07 & \cite{ols98}  & \nodata  \\
NGC1928   & $-$1.27 & 0.14 & 10.13 &  0.1  & \cite{mac04}  & {a}  \\
NGC1939   & $-$2.10 & 0.19 & 10.13 &  0.1  & \cite{mac04}  & {a}  \\
NGC1948   & $-$0.4  & 0.0  &  6.85 &  0.15 & \cite{wil96}  & {b}  \\
\enddata
\end{deluxetable*}

\tablenum{1}
\begin{deluxetable*}{rrrrrrl}
\tablecaption{LMC Cluster Ages and Metallicities (cont.)}
\tablewidth{0pt}
\tablehead{
  \colhead{Cluster} &
  \colhead{[Fe/H]} & \colhead{$\sigma_{[Fe/H]}$} &
  \colhead{log(Age)} & \colhead{$\sigma_{log(Age)}$} &
  \colhead{Reference} & \colhead{Notes}
}
\startdata
NGC1955   & $-$0.4  & 0.0  &  7.19 &  0.15 & \cite{dh98}   & {b}  \\
NGC1978   & $-$0.38 & 0.07 &  9.54 &  0.0  & \cite{fer06}  & {ce} \\
NGC2004   & $-$0.4  & 0.0  &  7.19 &  0.15 & \cite{dh98}   & {be} \\
NGC2005   & $-$1.35 & 0.15 & 10.22 &  0.11 & \cite{ols98}  & \nodata  \\
NGC2019   & $-$1.23 & 0.15 & 10.21 &  0.08 & \cite{ols98}  & \nodata  \\
NGC2027   & $-$0.4  & 0.0  &  7.06 &  0.14 & \cite{dh98}   & {b}  \\
NGC2031   & $-$0.52 & 0.21 &  8.2  &  0.1  & \cite{dir00}  & \nodata  \\
NGC2121   & $-$0.4  & 0.1  &  9.46 &  0.07 & \cite{ker07}  & {e}  \\
NGC2134   & $-$0.4  & 0.0  &  8.28 &  0.0  & \cite{val94a} & {bc} \\
NGC2136   & $-$0.55 & 0.23 &  8.0  &  0.1  & \cite{dir00}  & \nodata  \\
NGC2155   & $-$0.7  & 0.1  &  9.48 &  0.03 & \cite{ker07}  & {e}  \\
NGC2162   & $-$0.4  & 0.1  &  9.10 &  0.03 & \cite{ker07}  & \nodata  \\
NGC2164   & $-$0.2  & 0.2  &  8.0  &  0.0  & \cite{sr91}   & {c}  \\
NGC2173   & $-$0.6  & 0.1  &  9.21 &  0.04 & \cite{ker07}  & {e}  \\
NGC2209   & $-$0.5  & 0.1  &  9.08 &  0.03 & \cite{ker07}  & \nodata  \\
NGC2210   & $-$1.75 & 0.1  & 10.13 &  0.1  & \cite{hill00} & {ehi}\\           
NGC2213   & $-$0.7  & 0.1  &  9.23 &  0.04 & \cite{ker07}  & {e}  \\
NGC2214   & $-$0.2  & 0.2  &  8.0  &  0.0  & \cite{sr91}   & {c}  \\
NGC2249   & $-$0.45 & 0.1  &  9.35 &  0.03 & \cite{ker07}  & \nodata  \\
NGC2257   & $-$1.63 & 0.21 & 10.2  &  0.1  & \cite{dir00}  & \nodata  \\
\enddata

\tablenotetext{a}{age found to be ``equal to M92''; age error is our
  estimate from isochrones}
\tablenotetext{b}{no [Fe/H] error given}
\tablenotetext{c}{no age error given}
\tablenotetext{d}{age reference: \cite{joh99}}
\tablenotetext{e}{discrepant [Fe/H] measurements exist in literature}
\tablenotetext{f}{age reference: \cite{bic98}}
\tablenotetext{g}{age reference: \cite{ker07}}
\tablenotetext{h}{age found to be ``equal to M68''; age error is our
  estimate from isochrones}
\tablenotetext{i}{age reference: \cite{bro96}}

\end{deluxetable*}

\end{document}